\documentclass[twocolumn]{emulateapj}
\usepackage{ amsmath, color, natbib,graphicx,tablefootnote}

\newcommand\ii{{\sc ii}}
\newcommand\iii{{\sc iii}}

\shorttitle{CHAOS II}
\shortauthors{Croxall et al.}
\begin{document}  
\title{CHAOS II: Gas-Phase Abundances in NGC~5194}
\author{K.V. Croxall$^{1}$, R.W. Pogge$^{1,2}$,  D. Berg$^{3,4}$, E.~D. Skillman$^{4}$, J. Moustakas$^{5}$}
\affil{$^1$Department of Astronomy, The Ohio State University, 140 W 18th Ave., Columbus, OH, 43210
\\$^2$Center for Cosmology \& AstroParticle Physics, The Ohio State University, 191 West Woodruff Avenue, Columbus, OH 43210
\\$^3$Center for Gravitation, Cosmology and Astrophysics, Department of Physics, University of Wisconsin Milwaukee, 1900 East Kenwood Boulevard, Milwaukee, WI 53211, USA
\\$^4$Minnesota Institute for Astrophysics, University of Minnesota, 116 Church St. SE, Minneapolis, MN 55455 
\\$^5$Department of Physics \& Astronomy, Siena College, 515 Loudon Road, Loudonville, NY 12211
}

\email{croxall.5@osu.edu}

\begin{abstract}
We have observed NGC~5194 (M51a) as part of the CHemical Abundances of Spirals (CHAOS) project. Using the Multi Object Double Spectrographs (MODS) on the Large Binocular Telescope (LBT) we are able to measure one or more of the temperature-sensitive auroral lines ([O\,\iii] $\lambda$4363, [N\,\ii] $\lambda$5755, [S\,\iii] $\lambda$6312) and thus measure ``direct" gas-phase abundances in 29 individual H\ii\ regions.  [O\,\iii] $\lambda$4363 is only detected in two H\ii\ regions both of which show indications of excitation by shocks.  We compare our data to previous direct abundances measured in NGC\,5194 and find excellent agreement for all but one region ($\Delta[log(O/H)]\approx0.04$).  We find no evidence of trends in Ar/O, Ne/O, or S/O within NGC\,5194 or compared to other galaxies.  We find modest negative gradients in both O/H and N/O with very little scatter ($\sigma~\le~$0.08\,dex), most of which can be attributed to random error and not to intrinsic dispersion.  The gas-phase abundance gradient is consistent with the gradients observed in other interacting galaxies, which tend to be shallower than gradients measured in isolated galaxies.  The N/O ratio ($<$log(N/O)$>$ = $-$0.62) suggests secondary nitrogen production is responsible for a significantly larger fraction of nitrogen (e.g., factor of 8-10) relative to primary production mechanisms than predicted by theoretical models.
\end{abstract} 
 
 
 \section{Introduction}
In the present epoch, the majority of current star formation occurs in spiral galaxies. Spiral galaxies thus provide the best laboratories for understanding the star formation process and its effects upon the evolution of a galaxy.  In order to understand the chemical evolution and nucleosynthesis that has occurred in spiral galaxies we must first measure the bulk abundance of metals in these galaxies.  While seemingly straightforward, it is made challenging by the innate difficulty of measuring the electron temperatures (T$_e$) of H\ii\ regions which enable a direct determination of gas-phase abundances.  This difficulty is particularly noticeable in metal-rich massive spiral galaxies where temperature-sensitive auroral lines are exceptionally weak \citep{stasinska2005}. 

Notwithstanding the difficulties, metal rich galaxies play a central role in the creation of heavy elements.   High quality observations of extragalactic H\ii\ regions located in massive spirals, i.e., metal rich environments, have revealed significant discrepancies between the indirect measures of the gas-phase metallicity and oxygen abundances based on direct measurements of T$_e$ \citep[e.g.,][]{Kinkel1994, pilyugin2010}.  In part, this discrepancy may be attributed to fact that direct abundances have been predominantly observed in metal poor environments where the [O\,\iii] $\lambda$4363 auroral line can be more easily detected.  This observational bias leads to a less reliable calibration of indirect strong line abundance calibrations as the data in this region of parameter space are sparsely sampled and biased towards lower-metallicity regions \citep{moustakas2010}.

Furthermore, the paucity of data renders us unable to deduce the detailed chemical structure of massive galaxies.  Indeed, sparsely sampled maps of gas-phase abundances are unable to address the amount of dispersion seen in metallicity at a given radius and may give an inaccurate representative abundance at a given radius as their slopes may be predominantly set by one or two H\ii\ regions which act as lever arms \citep[e.g.,][]{bresolin2004, berg2013}.  

Given (1) the inherent faintness of the auroral lines necessary to determine an electron temperature, particularly at high metallicities where the far-IR lines begin to dominate the cooling of H\ii\ regions, and (2) the need for large numbers of H\ii\ regions with robust measurements of T$_e$, we have undertaken a large spectroscopic study to definitively measure the CHemical Abundances Of Spirals \citep[CHAOS, see][]{berg2014}.  While the galaxies selected for this study are well studied galaxies \citep[see][]{sings, kingfish}, robust, high signal-to-noise spectroscopic observations of H\ii\ regions spanning their disks are still lacking for many \citep[see][]{moustakas2010}.  In this work we present spectral observations of the nearby spiral NGC~5194 (also known as M51a, the Whirlpool Galaxy).

As a massive interacting spiral galaxy \citep[log(M$_{\star})\sim$10.5,][]{leroy2013}, NGC~5194 has numerous H\ii\ regions with a relatively high oxygen abundance \citep[12 + log(O/H) $\approx$ 8.8,][]{garnett2004,bresolin2004}.  In the expected range of oxygen abundances, the nominal temperature-sensitive line [O~\iii] $\lambda$4363 is not expected to be detected.  Instead, the auroral [N\,\ii] $\lambda$5755 line which is more accessible in cooler, low-excitation H\ii\ regions can be observed.  \citet{garnett2004} and \citet{bresolin2004} showed this was indeed the case by detecting the [N\,\ii] $\lambda$5755 line in a total of 10 H\ii\ regions in NGC~5194.  However, they found a metallicity that was roughly a factor of three smaller than expected, an elevated N/O ratio compared to similar galaxies, and a shallow gradient compared to other large spirals \citep{bresolin2004}.  

Given the implications for abundance gradients and strong-line calibrations, we re-examine this quintessential metal-rich galaxy.  In particular we seek to better determine the abundance gradient.  The slope found by \citet{bresolin2004} relies primarily on two lever points: one in the interior of the galaxy (R/R$_{25}\sim$\,0.19 ) and one in the outskirts (R/R$_{25}\sim$\,1.04).  Accordingly, we have observed H\ii\ regions in NGC~5194 as part of CHAOS to increase the number of auroral line detections, verify its absolute metallicity and metallicity gradient, and investigate multiple means of determining the electron temperature in metal-rich galaxies.  Our observations and and data reduction are described in \S2.  In \S3 we determine electron temperatures and direct gas-phase chemical abundances.  We present abundance gradients for O/H and N/O in \S4.  We discuss these abundance gradients and the implications on chemical evolution in \S5.   Finally, we summarize our conclusions in~\S6.

\section {Observations}
\subsection{Optical Spectroscopy}
Optical spectra of NGC\,5194 were acquired with the Multi-Object Double Spectrographs \citep[MODS,][]{mods} on the Large Binocular Telescope (LBT) as part of the CHAOS study during April 2012.  At the time of the observations, both spectrographs were not available.  Thus, we acquired all spectra of NGC~5194 using MODS1.  We obtained simultaneous blue and red spectra using the G400L (400 lines mm$^{-1}$, R$\approx$1850) and G670L (250 lines mm$^{-1}$, R$\approx$2300) gratings, respectively.  This setup provided broad spectral coverage extending from 3100 -- 10,000 \AA.  In order to detect the intrinsically weak auroral lines, i.e., [N\,\ii] $\lambda$5755 and [S\,\iii] $\lambda$6312, in numerous H\ii\ regions, three fields, each containing $\sim$20 slits, were targeted with six 1200s exposures, for a total integration time of 2-hours for each field.  Given the expected high metallicity of NGC~5194, and hence cool temperatures, we did not expect to detect the [O~\iii] $\lambda$4363 auroral line using these exposure times.  

\begin{figure}[bp] 
\epsscale{1.2}
   \centering
   \plotone{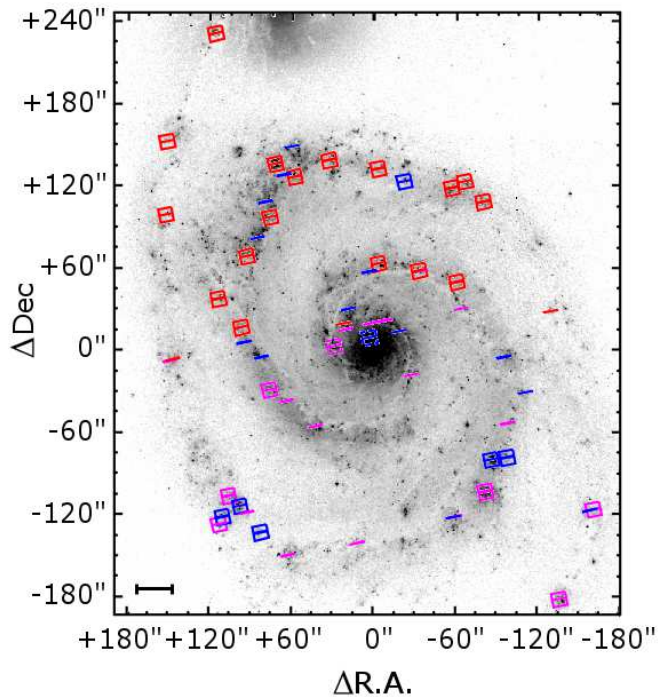}
   \caption{Map of targeted H\ii\ regions in NGC~5194 overlaid on an HST H$\alpha$ image.  Offsets in arc seconds are relative to the galaxy center given in Table 1.  Slits targeting H\ii\ regions in which we were able to measure at least one temperature-sensitive auroral line are circumscribed by squares.  Field 1 slits are shown in red; Field 2 slits are shown in blue; Field 3 slits are shown in magenta.  Dashed lines indicate regions in which indications of significant shock ionization are present. A scale bar showing the size of 1000\,pc is shown in the lower left.}
   \label{fig:slits}
\end{figure}

Figure \ref{fig:slits} shows the locations of slits for each of the three observed MODS fields.  Throughout this work, all locations are listed as offsets, in right ascension and declination, from the center of NGC~5194, as listed in Table \ref{t:m51global}.  Slits were cut to lie close to the median parallactic angle of the observing window in order to minimize light loss due to atmospheric refraction.  Our targeted regions (see Table \ref{t:locations}), as well as alignment stars, were selected based on archival broad-band and H$\alpha$ imaging from the Hubble Space Telescope\footnote{HST Project 10452, Cycle 13}.  We cut most slits to be 10\arcsec\ long with a 1\arcsec\ slit width.  Slits were placed on relatively bright H\ii\ regions across the entirety of the disk; this procedure ensured that both radial and azimuthal  trends in the abundances could be studied.  When extra space between slits was available, slits were extended to make the best use of the available field of view.  Line emission was detected in all 61 slits cut in our masks.

\begin{deluxetable}{lcccccccccc}  
\tabletypesize{\scriptsize}
\tablecaption{Adopted Global Properties of NGC 5194}
\tablewidth{0pt}
\tablehead{ 
  \colhead{Property}	&
  \colhead{Adopted Value}	&
  \colhead{Reference}	
  }
\startdata
R.A. 	& 13$^h$29$^m$52.7$^s$ &  1 \\
Dec  & +47$^\circ$11$^m$43$^s$ & 1 \\
Inclination & 22$^\circ$ & 2\\
Position Angle  & 172$^\circ$ & 3\\
Distance & 7.9 Mpc & 1 \\
R$_{25}$ & 336.6\arcsec & 4 
\enddata
\label{t:m51global}
\tablecomments{Units of right ascension are hours, minutes, and seconds, and units of declination are degrees, arcminutes, and arcseconds. References are as follows: [1]  NED [2]  Colombo et al. 2014 [3] Walter et al. 2008 [4] RC3}
\end{deluxetable}

\begin{deluxetable*}{lccccc|c|c|c|c|c|c|c}  
	\tabletypesize{\scriptsize}
	\tablecaption{NGC 5194 MODS/LBT Observations}
	\tablewidth{0pt}
	\tablehead{   
	  \colhead{H~\ii}	&
	  \colhead{R.A.}		&
	  \colhead{Dec.}	&
	  \colhead{R$_g$}	&
	  \colhead{R/R$_{25}$}	&
	  \colhead{R$_g$}	&
	  \multicolumn{5}{c}{Auroral Line Detections} &
	  \colhead{Wolf}\\
	  \colhead{Region}	&
	  \colhead{(2000)}	&
	  \colhead{(2000)}	&
	  \colhead{(arcsec)}	&
	  \colhead{}		&
	  \colhead{(kpc)}	&
	  \colhead{[O \iii]}	&
	  \colhead{[N \ii]}	&
	  \colhead{[S \iii]}	&
	  \colhead{[O \ii]}	&
	  \colhead{[S \ii]}	&
	  \colhead{Rayet}	
	  }
\startdata
Total Detections: & & & & & & 2 & 28 & 17 & 27 & 18 & 11\\\hline
NGC5194+2.5+9.5 	&	13:29:52.9	&	47:11:52.49	&	10.26	&	0.030	&	0.32	&	\checkmark	&	\checkmark	&		&	\checkmark	&	\checkmark	&		\\
NGC5194+1.1+19.8 	&	13:29:52.8	&	47:12:02.76	&	20.23	&	0.060	&	0.64	&		&		&		&		&		&		\\
NGC5194+0.2+20.2 	&	13:29:52.7	&	47:12:03.18	&	20.59	&	0.061	&	0.65	&		&		&		&		&		&		\\
NGC5194-6.9+20.8 	&	13:29:52.0	&	47:12:03.76	&	22.35	&	0.066	&	0.70	&		&		&		&		&		&		\\
NGC5194-18.5+13.7 	&	13:29:50.9	&	47:11:56.65	&	24.29	&	0.072	&	0.77	&		&		&		&		&		&		\\
NGC5194+19.7+15.0 	&	13:29:54.6	&	47:11:58.04	&	26.38	&	0.078	&	0.83	&		&		&		&		&		&		\\
NGC5194+27.2+3.6 	&	13:29:55.4	&	47:11:46.57	&	29.54	&	0.088	&	0.93	&		&		&		&		&		&		\\
NGC5194+21.3+19.6 	&	13:29:54.8	&	47:12:02.64	&	30.70	&	0.091	&	0.97	&		&		&		&		&		&		\\
NGC5194+30.2+2.2 	&	13:29:55.7	&	47:11:45.24	&	32.56	&	0.097	&	1.03	&	\checkmark	&	\checkmark	&		&	\checkmark	&	\checkmark	&	\checkmark	\\
NGC5194-27.9-18.1 	&	13:29:50.0	&	47:11:24.91	&	35.33	&	0.105	&	1.11	&		&		&		&		&		&		\\
NGC5194+16.3+30.8 	&	13:29:54.3	&	47:12:13.81	&	36.09	&	0.107	&	1.14	&		&		&		&		&		&		\\
NGC5194+17.7+30.2 	&	13:29:54.4	&	47:12:13.20	&	36.32	&	0.108	&	1.14	&		&		&		&		&		&		\\
NGC5194+4.1+56.5 	&	13:29:53.1	&	47:12:39.54	&	57.26	&	0.170	&	1.80	&		&		&		&		&		&		\\
NGC5194-4.3+63.3 	&	13:29:52.3	&	47:12:46.34	&	63.91	&	0.190	&	2.01	&		&	\checkmark	&		&	\checkmark	&		&	\checkmark	\\
NGC5194-33.2+58.0 	&	13:29:49.4	&	47:12:40.96	&	67.95	&	0.202	&	2.14	&		&	\checkmark	&		&	\checkmark	&	\checkmark	&	\checkmark	\\
NGC5194+42.7-55.8 	&	13:29:56.9	&	47:10:47.18	&	71.26	&	0.212	&	2.25	&		&		&		&		&		&		\\
NGC5194+62.7-36.9 	&	13:29:58.8	&	47:11:06.11	&	75.99	&	0.226	&	2.39	&		&		&		&		&		&		\\
NGC5194-65.0+30.5 	&	13:29:46.3	&	47:12:13.52	&	76.12	&	0.226	&	2.40	&		&		&		&		&		&		\\
NGC5194-62.2+50.3 	&	13:29:46.6	&	47:12:33.26	&	83.28	&	0.247	&	2.62	&		&	\checkmark	&		&	\checkmark	&		&		\\
NGC5194+75.5-28.7 	&	13:30:00.1	&	47:11:14.29	&	85.39	&	0.254	&	2.69	&		&		&	\checkmark	&	\checkmark	&		&		\\
NGC5194+81.9-5.4 	&	13:30:00.7	&	47:11:37.54	&	88.17	&	0.262	&	2.78	&		&		&		&		&		&		\\
NGC5194+93.6+5.9 	&	13:30:01.9	&	47:11:48.91	&	100.95	&	0.300	&	3.18	&		&		&		&		&		&		\\
NGC5194-96.5-4.5 	&	13:29:43.2	&	47:11:38.49	&	104.25	&	0.310	&	3.29	&		&		&		&		&		&		\\
NGC5194+96.1+16.8 	&	13:30:02.1	&	47:11:59.77	&	105.13	&	0.312	&	3.31	&		&	\checkmark	&		&	\checkmark	&		&		\\
NGC5194-98.9-52.8 	&	13:29:43.0	&	47:10:50.18	&	119.88	&	0.356	&	3.78	&		&		&		&		&		&		\\
NGC5194+91.0+69.0 	&	13:30:01.6	&	47:12:52.00	&	121.24	&	0.360	&	3.82	&		&	\checkmark	&	\checkmark	&	\checkmark	&		&		\\
NGC5194+83.6+82.0 	&	13:30:00.9	&	47:13:05.02	&	123.35	&	0.366	&	3.89	&		&		&		&		&		&		\\
NGC5194-110.3-31.3 	&	13:29:41.9	&	47:11:11.68	&	123.56	&	0.367	&	3.89	&		&		&		&		&		&		\\
NGC5194-86.5-79.4 	&	13:29:44.2	&	47:10:23.58	&	123.57	&	0.367	&	3.89	&		&	\checkmark	&	\checkmark	&	\checkmark	&		&	\checkmark	\\
NGC5194-22.5+122.8 	&	13:29:50.5	&	47:13:45.84	&	125.29	&	0.372	&	3.95	&		&	\checkmark	&		&		&		&		\\
NGC5194+112.7+37.7 	&	13:30:03.8	&	47:12:20.62	&	127.83	&	0.380	&	4.03	&		&	\checkmark	&		&		&	\checkmark	&		\\
NGC5194+76.6+96.3 	&	13:30:00.2	&	47:13:19.25	&	128.39	&	0.381	&	4.05	&		&	\checkmark	&		&	\checkmark	&	\checkmark	&		\\
NGC5194-97.0-78.4 	&	13:29:43.2	&	47:10:24.55	&	131.92	&	0.392	&	4.16	&		&		&	\checkmark	&	\checkmark	&		&		\\
NGC5194-3.0+131.9 	&	13:29:52.4	&	47:13:54.85	&	132.42	&	0.393	&	4.17	&		&	\checkmark	&	\checkmark	&	\checkmark	&	\checkmark	&	\checkmark	\\
NGC5194-57.2+118.2 	&	13:29:47.1	&	47:13:41.22	&	132.75	&	0.394	&	4.18	&		&	\checkmark	&	\checkmark	&	\checkmark	&	\checkmark	&		\\
NGC5194-78.9+107.4 	&	13:29:45.0	&	47:13:30.42	&	136.07	&	0.404	&	4.29	&		&	\checkmark	&		&		&	\checkmark	&		\\
NGC5194-82.0-102.7 	&	13:29:44.6	&	47:10:00.24	&	136.82	&	0.406	&	4.31	&		&	\checkmark	&		&	\checkmark	&		&		\\
NGC5194-59.0-121.4 	&	13:29:46.9	&	47:09:41.63	&	138.07	&	0.410	&	4.35	&		&		&		&		&		&		\\
NGC5194-60.0-121.3 	&	13:29:46.8	&	47:09:41.65	&	138.54	&	0.412	&	4.37	&		&		&		&		&		&		\\
NGC5194+77.5+108.4 	&	13:30:00.3	&	47:13:31.43	&	138.59	&	0.412	&	4.37	&		&		&		&		&		&		\\
NGC5194+12.0-140.7 	&	13:29:53.9	&	47:09:22.28	&	140.87	&	0.419	&	4.44	&		&		&		&		&		&		\\
NGC5194-66.6+122.9 	&	13:29:46.2	&	47:13:45.87	&	141.56	&	0.421	&	4.46	&		&	\checkmark	&	\checkmark	&	\checkmark	&	\checkmark	&	\checkmark	\\
NGC5194+13.3-141.3 	&	13:29:54.0	&	47:09:21.69	&	141.56	&	0.421	&	4.46	&		&		&		&		&		&		\\
NGC5194-129.7+28.3 	&	13:29:40.0	&	47:12:11.25	&	142.14	&	0.422	&	4.48	&		&		&		&		&		&		\\
NGC5194+56.8+126.5 	&	13:29:58.3	&	47:13:49.52	&	142.17	&	0.422	&	4.48	&		&	\checkmark	&		&	\checkmark	&		&		\\
NGC5194+30.8+139.0 	&	13:29:55.7	&	47:14:02.03	&	144.16	&	0.428	&	4.54	&		&	\checkmark	&	\checkmark	&	\checkmark	&	\checkmark	&	\checkmark	\\
NGC5194+65.0+127.7 	&	13:29:59.1	&	47:13:50.71	&	147.34	&	0.438	&	4.64	&		&		&		&		&		&		\\
NGC5194+104.1-105.5 	&	13:30:02.9	&	47:09:57.48	&	152.09	&	0.452	&	4.79	&		&	\checkmark	&	\checkmark	&	\checkmark	&		&	\checkmark	\\
NGC5194+92.3-118.5 	&	13:30:01.7	&	47:09:44.50	&	152.83	&	0.454	&	4.82	&		&		&		&		&		&		\\
NGC5194+98.1-113.8 	&	13:30:02.3	&	47:09:49.18	&	153.39	&	0.456	&	4.83	&		&	\checkmark	&	\checkmark	&	\checkmark	&	\checkmark	&	\checkmark	\\
NGC5194+146.4-6.7 	&	13:30:07.1	&	47:11:36.29	&	157.51	&	0.468	&	4.96	&		&		&		&		&		&		\\
NGC5194+71.2+135.9 	&	13:29:59.7	&	47:13:58.87	&	157.89	&	0.469	&	4.98	&		&	\checkmark	&	\checkmark	&	\checkmark	&		&	\checkmark	\\
NGC5194+83.4-133.1 	&	13:30:00.9	&	47:09:29.86	&	158.81	&	0.472	&	5.00	&		&	\checkmark	&	\checkmark	&	\checkmark	&	\checkmark	&	\checkmark	\\
NGC5194+147.7-7.6 	&	13:30:07.2	&	47:11:35.33	&	158.94	&	0.472	&	5.01	&		&		&		&		&		&		\\
NGC5194+63.1-149.7 	&	13:29:58.9	&	47:09:13.33	&	162.88	&	0.484	&	5.13	&		&		&		&		&		&		\\
NGC5194+60.2+147.9 	&	13:29:58.6	&	47:14:10.86	&	163.18	&	0.485	&	5.14	&		&		&		&		&		&		\\
NGC5194+109.9-121.4 	&	13:30:03.5	&	47:09:41.59	&	167.56	&	0.498	&	5.28	&		&	\checkmark	&	\checkmark	&	\checkmark	&	\checkmark	&		\\
NGC5194+112.2-126.6 	&	13:30:03.7	&	47:09:36.40	&	172.95	&	0.514	&	5.45	&		&	\checkmark	&		&		&	\checkmark	&		\\
NGC5194+150.6+99.0 	&	13:30:07.5	&	47:13:21.97	&	192.00	&	0.570	&	6.05	&		&	\checkmark	&	\checkmark	&	\checkmark	&	\checkmark	&		\\
NGC5194-159.5-116.4 	&	13:29:37.0	&	47:09:46.50	&	209.56	&	0.623	&	6.60	&		&	\checkmark	&	\checkmark	&	\checkmark	&	\checkmark	&		\\
NGC5194+150.3+152.1 	&	13:30:07.5	&	47:14:15.07	&	224.77	&	0.668	&	7.08	&		&		&		&	\checkmark	&	\checkmark	&		\\
NGC5194-135.4-181.4 	&	13:29:39.4	&	47:08:41.50	&	235.20	&	0.699	&	7.41	&		&	\checkmark	&	\checkmark	&	\checkmark	&	\checkmark	&		\\
NGC5194+114.5+230.8 	&	13:30:03.9	&	47:15:33.79	&	264.57	&	0.786	&	8.34	&		&	\checkmark	&	\checkmark	&	\checkmark	&		&		
\enddata
	\label{t:locations}
	\tablecomments{Observing logs for H~\ii\ regions observed in NGC 5194 using MODS on the LBT on the UT dates of April 29 and 30, 2012. Each field was observed over an integrated exposure time of 1200s on clear nights, with, on average $\sim$1\farcs00 seeing, and airmasses less than 1.5. Slit ID, composed of the galaxy name and the offset in R.A. and Dec., in arcseconds, from the central position listed in Table \ref{t:m51global} is listed in Column 1. The right ascension and declination of the individual H II regions are given in units of hours, minutes, and seconds, and degrees, arcminutes, and arcseconds respectively in columns 2 and 3.  The de-projected distances of H~\ii\ regions from the center of the galaxy in arcseconds, fraction of R$_{25}$, and in kpc are listed in the Columns 4-6.  Columns 7-11 highlight which regions have [O~\iii] $\lambda$4363, [N\,\ii] $\lambda$5755, [S\,\iii] $\lambda$6312, [O~\ii] 7330 auroral lines detections at the 3$\sigma$ significance level.  Finally, columns 12 and 13 indicate which H\ii\ regions have detections of the nebular Balmer continuum jump and Wolf-Rayet features.}
\end{deluxetable*}

\begin{figure*}[tbp] 
\epsscale{1.0}
   \centering
   \plotone{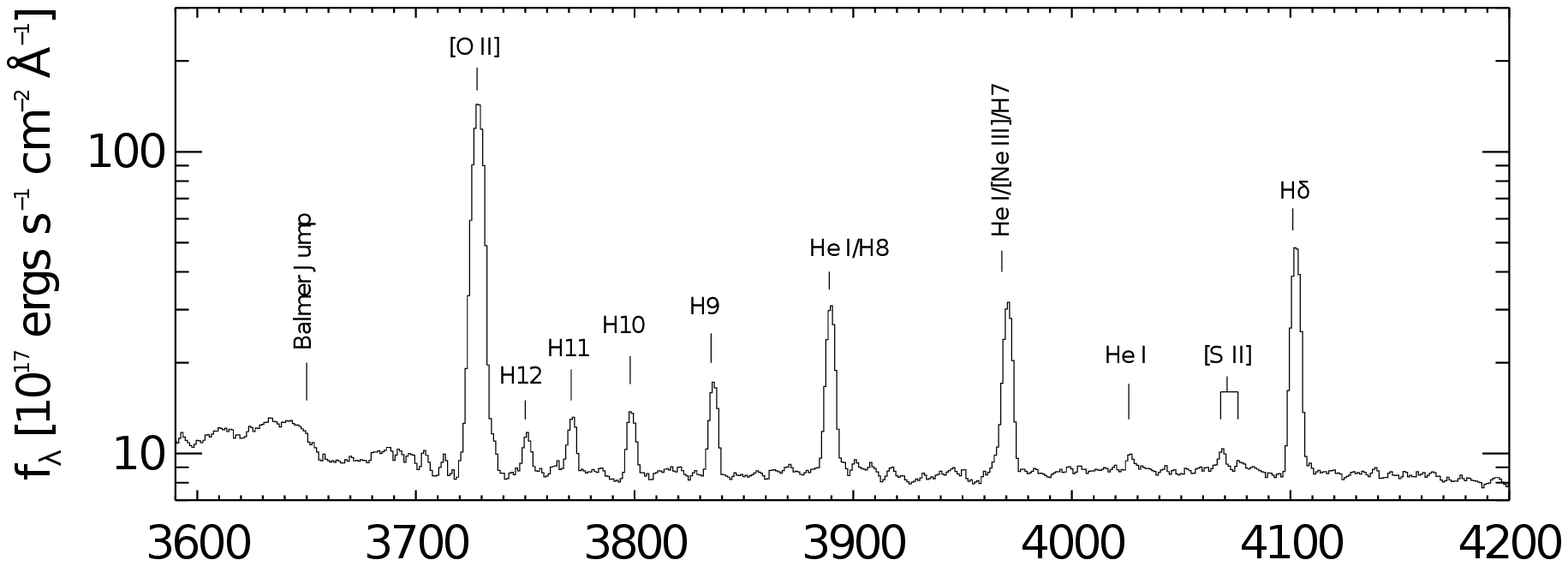}
   \plotone{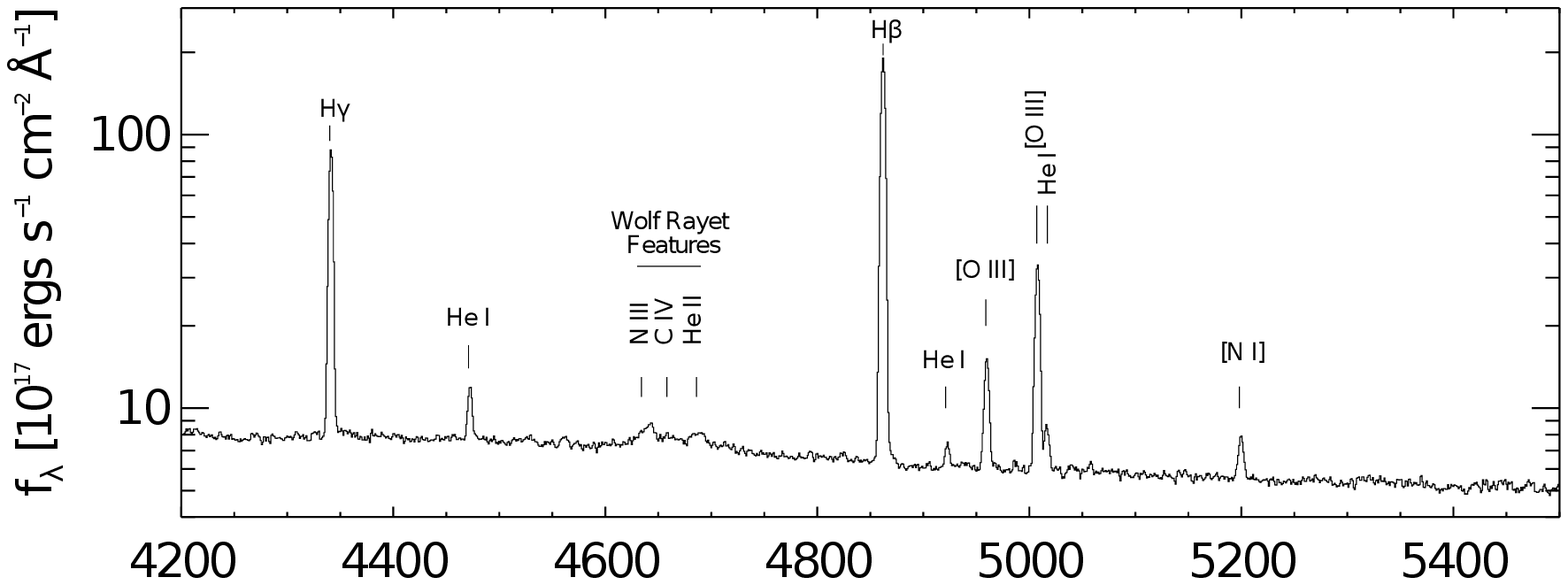}
   \plotone{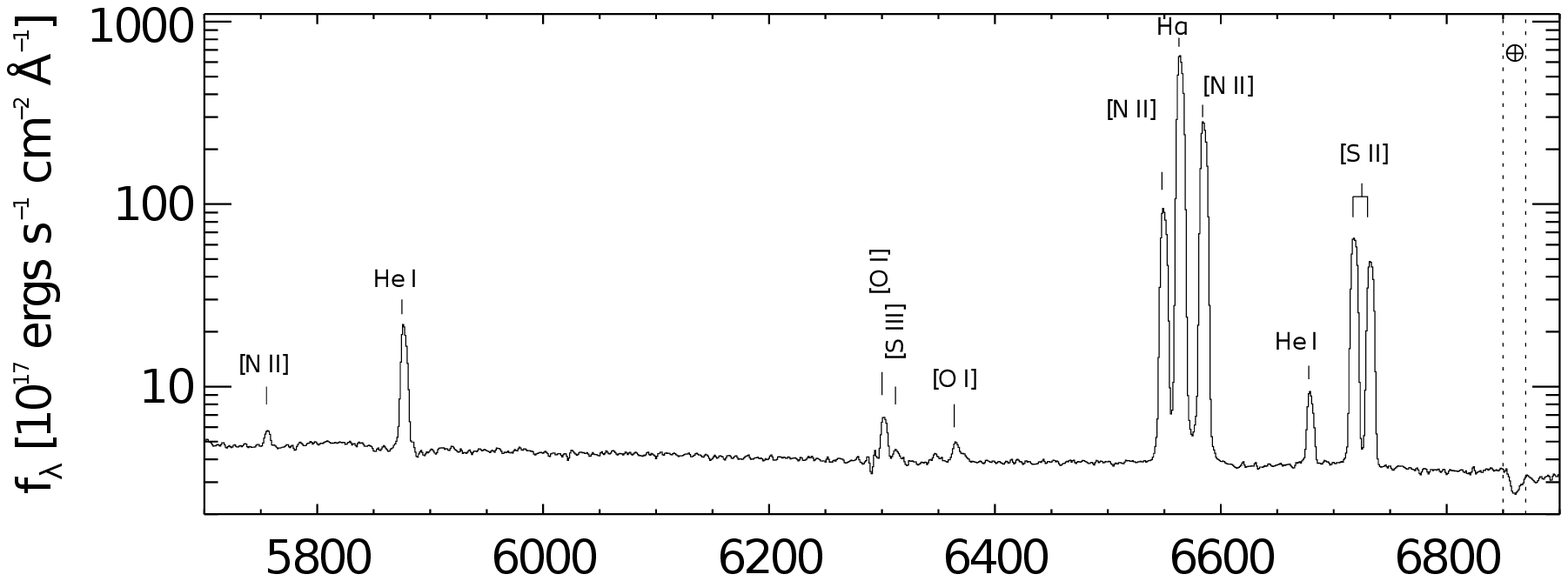}
   \plotone{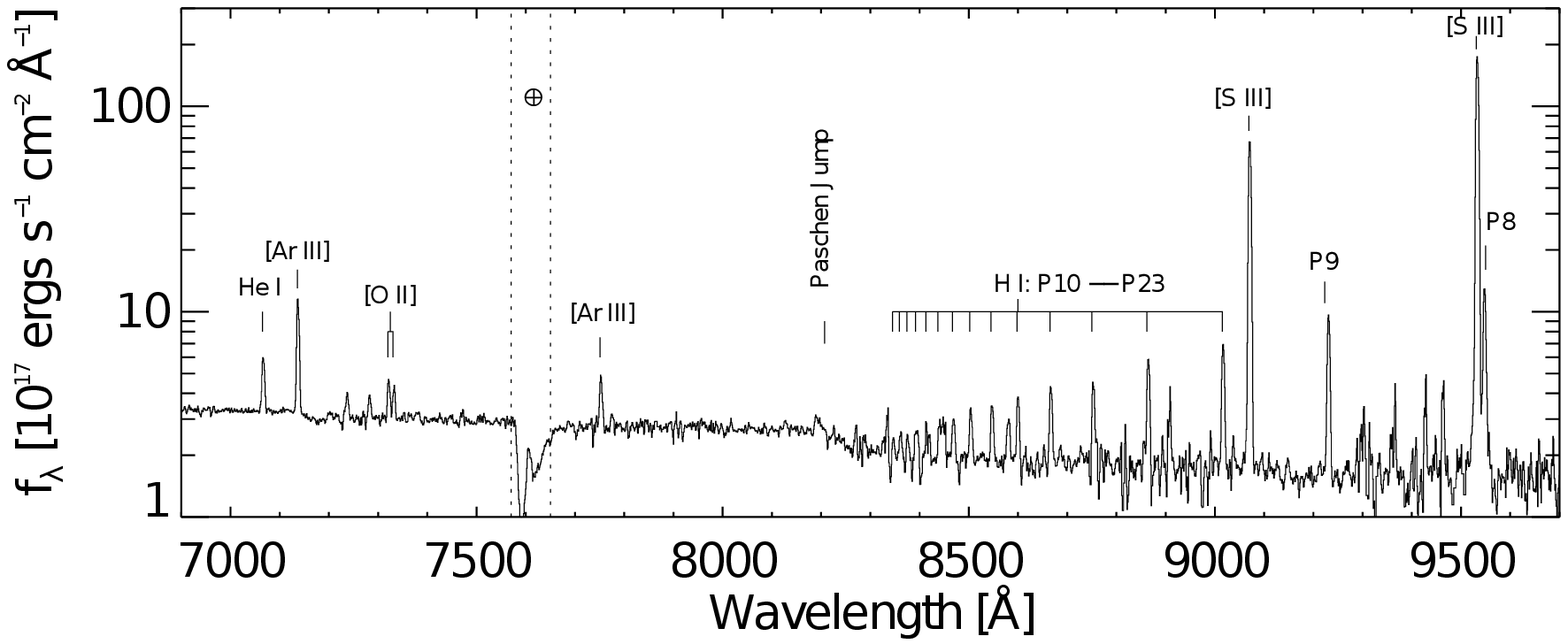}
   \caption{Example of a one dimensional spectrum taken from MODS1 observations of NGC~5194.  Notable major emission features are marked and labeled.  The [N\,\ii] $\lambda$5755 line in this spectrum is characterized by a signal-to-noise of $\sim$12, with a peak that is $\sim$67 greater than the RMS noise. We have not corrected for major telluric absorption features, which are also marked in the spectrum.} 
   \label{fig:spectra}
\end{figure*}

For a detailed description of the data reduction procedures we refer the reader to \citet{berg2014}.  Here we will only note the key points of our data processing.  Spectra were reduced and analyzed using the Beta-version of the MODS reduction pipeline\footnote{http://www.astronomy.ohio-state.edu/MODS/Software/modsIDL/} which runs within the XIDL\footnote{http://www.ucolick.org/$\sim$xavier/IDL/} reduction package.  As no sky-only slits were cut in these masks, a two-dimensional sky frame was created by fitting a two-dimensional B-spline to the background sky in the two-dimensional spectra, i.e., each slit was fit independently with a local sky.  This sky model was compared to the model for the two slits with the largest sections of clean background sky-emission.  When oversubtraction of strong lines was un-avoidable using a local sky, the representative sky spectrum was used. One-dimensional spectra were corrected for atmospheric extinction and flux calibrated based on observations of flux standard stars \citep{Bohlin2010}. At least one flux standard was observed on each night science data were obtained. An example of a flux-calibrated spectrum is shown in Figure \ref{fig:spectra}. 
 
\subsection{Spectral Modeling and Line Intensities}
We provide detailed descriptions of the adopted continuum modeling and line fitting procedures applied to the CHAOS observations in \citet{berg2014}.  In this paper we will only highlight the fundamental components of this process. 

We model the underlying continuum of our MODS1 spectra using the STARLIGHT\footnote{www.starlight.ufsc.br} spectral synthesis code \citep{starlight} using the models of \citet{bruzual}. Allowing for an additional faint nebular continuum we fit each emission line with a Gaussian profile.  We note that we have modeled blended lines (H7, H8, and H11 -- H14) in the Balmer series based on the measurements of unblended Balmer lines and the tabulated atomic ratios of \citet{hummer}, assuming Case B recombination.  As the [N\,\ii] $\lambda$5755 line is vital to this study and lies in the region where the MODS dichroic can distort the shape of the continuum, we measured the flux of this line by hand in each extracted spectrum rather than adopting the automated fit to this line.

We correct the strength of emission features for line-of-sight reddening using the relative intensities of the three strongest Balmer lines (H$\alpha$/H$\beta$, H$\alpha$/H$\gamma$, H$\beta$/H$\gamma$).  We report the determined values of c(H$\beta$) in Table \ref{t:lineflux}.  We do not apply an ad-hoc correction to account for Balmer absorption as the lines were fit simultaneously with the stellar population models.  

\begin{deluxetable*}{lccccccccccccccccccccccccccccccccccccccccccccccccccccccccccccccccccc}  
	\tabletypesize{\scriptsize}
	\tablecaption{Line fluxes relative to H$\beta$}
	\tablewidth{0pt}
	\tablehead{   
	  \colhead{Ion}	&
	  \colhead{NGC5194+3.3+9.1 }		&
	  \colhead{NGC5194+2.5+9.5 }		&
	  \colhead{NGC5194+1.1+19.8 }	&
	  \colhead{NGC5194+0.2+20.2 }	&
	  \colhead{NGC5194-6.9+20.8 }		
}
\startdata
$[$O~\ii]~$\lambda$3726		&	$\leq$~0.0117&	2.44$\pm$0.40&	0.160$\pm$0.015&	0.200$\pm$0.021&	0.880$\pm$0.055\\
$[$O~\ii]~$\lambda$3728		&	2.2779$\pm$0.9031&	3.40$\pm$0.50&	0.150$\pm$0.016&	0.130$\pm$0.021&	$\leq$~0.0094\\
H13 $\lambda$3734		&	0.0062$\pm$0.0183&	0.0052$\pm$0.0064&	0.0243$\pm$0.0046&	0.0224$\pm$0.0062&	0.0086$\pm$0.0085\\
H12 $\lambda$3750		&	$\leq$~0.130&	$\leq$~0.0072&	$\leq$~0.036&	$\leq$~0.048&	$\leq$~0.0089\\
H11 $\lambda$3770		&	$\leq$~0.130&	$\leq$~0.0065&	0.062$\pm$0.010&	0.082$\pm$0.016&	$\leq$~0.072\\
H10 $\lambda$3797		&	$\leq$~0.120&	$\leq$~0.041&	0.0525$\pm$0.0098&	0.047$\pm$0.013&	$\leq$~0.055\\
He~I~$\lambda$3819		&	$\leq$~0.0102&	$\leq$~0.044&	$\leq$~0.0247&	$\leq$~0.040&	$\leq$~0.057\\
H9 $\lambda$3835		&	$\leq$~0.140&	0.0455$\pm$0.0153&	0.080$\pm$0.010&	0.120$\pm$0.015&	$\leq$~0.061\\
$[$Ne~\iii]~$\lambda$3868		&	$\leq$~0.120&	0.410$\pm$0.061&	$\leq$~0.042&	$\leq$~0.110&	0.200$\pm$0.022\\
He~I~$\lambda$3888		&	$\leq$~0.170&	0.170$\pm$0.031&	0.0682$\pm$0.0092&	0.120$\pm$0.015&	0.098$\pm$0.021\\
H8 $\lambda$3889		&	0.0252$\pm$0.0750&	0.0214$\pm$0.0263&	0.100$\pm$0.019&	0.089$\pm$0.025&	0.0353$\pm$0.0350\\
He~I~$\lambda$3964		&	$\leq$~0.110&	0.100$\pm$0.022&	$\leq$~0.0046&	$\leq$~0.00179&	$\leq$~0.0068\\
$[$Ne~\iii]~$\lambda$3967		&	0.1326$\pm$0.0617&	0.190$\pm$0.031&	$\leq$~0.0046&	0.120$\pm$0.013&	$\leq$~0.0067\\
H7 $\lambda$3970		&	0.0370$\pm$0.1102&	0.0317$\pm$0.0388&	0.150$\pm$0.028&	0.130$\pm$0.036&	0.0519$\pm$0.0516\\
$[$Ne~\iii]~$\lambda$4011		&	$\leq$~0.00277&	$\leq$~0.0038&	$\leq$~0.0033&	$\leq$~0.00118&	0.0356$\pm$0.0077\\
He~I~$\lambda$4026		&	$\leq$~0.00278&	$\leq$~0.0038&	$\leq$~0.0033&	$\leq$~0.00103&	$\leq$~0.0218\\
$[$S\,\ii]~$\lambda$4068		&	$\leq$~0.0297&	0.072$\pm$0.012&	0.0215$\pm$0.0057&	$\leq$~0.0181&	0.0305$\pm$0.0069\\
$[$S\,\ii]~$\lambda$4076		&	$\leq$~0.00249&	$\leq$~0.0210&	$\leq$~0.0152&	$\leq$~0.0160&	0.0331$\pm$0.0068\\
H$\delta$ $\lambda$4101		&	0.0638$\pm$0.0256&	0.120$\pm$0.017&	0.230$\pm$0.012&	0.290$\pm$0.015&	0.250$\pm$0.013\\
He~I~$\lambda$4120		&	$\leq$~0.00235&	$\leq$~0.00292&	$\leq$~0.00282&	$\leq$~0.0166&	$\leq$~0.0181\\
He~I~$\lambda$4143		&	$\leq$~0.00269&	$\leq$~0.00287&	$\leq$~0.00276&	$\leq$~0.00098&	$\leq$~0.0169\\
H$\gamma$ $\lambda$4340	&	0.270$\pm$0.087&	0.330$\pm$0.043&	0.460$\pm$0.021&	0.530$\pm$0.027&	0.490$\pm$0.023\\
$[$O~\iii]~$\lambda$4363		&	$\leq$~0.00228&	0.0478$\pm$0.0085&	$\leq$~0.0134&	$\leq$~0.0125&	$\leq$~0.0244\\
He~I~$\lambda$4387		&	$\leq$~0.00223&	$\leq$~0.00277&	$\leq$~0.0115&	$\leq$~0.00095&	$\leq$~0.00269\\
He~I~$\lambda$4471		&	$\leq$~0.00254&	0.0332$\pm$0.0073&	$\leq$~0.00260&	$\leq$~0.0128&	$\leq$~0.0144\\
$[$Fe~\iii]~$\lambda$4658		&	$\leq$~0.0231&	0.0217$\pm$0.0055&	0.0137$\pm$0.0036&	0.0151$\pm$0.0033&	$\leq$~0.00259\\
He~II~$\lambda$4685		&	$\leq$~0.00205&	0.0386$\pm$0.0065&	$\leq$~0.0098&	$\leq$~0.0103&	$\leq$~0.00247\\
H$\beta$ $\lambda$4861		&	1.00$\pm$0.30&	1.00$\pm$0.10&	1.000$\pm$0.043&	1.000$\pm$0.047&	1.000$\pm$0.044\\
He~I~$\lambda$4921		&	$\leq$~0.0050&	$\leq$~0.0047&	$\leq$~0.0035&	$\leq$~0.0128&	$\leq$~0.0039\\
$[$O~\iii]~$\lambda$4958		&	0.260$\pm$0.073&	0.590$\pm$0.066&	0.0648$\pm$0.0053&	0.0450$\pm$0.0043&	0.1290$\pm$0.0080\\
$[$O~\iii]~$\lambda$5006		&	0.50$\pm$0.10&	1.46$\pm$0.20&	0.1530$\pm$0.0083&	0.1110$\pm$0.0066&	0.360$\pm$0.017\\
He~I~$\lambda$5015		&	$\leq$~0.044&	$\leq$~0.0207&	$\leq$~0.0155&	$\leq$~0.0124&	0.0283$\pm$0.0063\\
NI $\lambda$5197		&	$\leq$~0.038&	0.280$\pm$0.031&	0.0230$\pm$0.0042&	0.0323$\pm$0.0041&	0.0764$\pm$0.0064\\
O~I~$\lambda$5577		&	$\leq$~0.046&	$\leq$~0.0262&	$\leq$~0.0241&	$\leq$~0.00274&	$\leq$~0.0077\\
$[$N\,\ii]~$\lambda$5754		&	$\leq$~0.0186&	0.0679$\pm$0.0074&	$\leq$~0.0070&	$\leq$~0.00093&	$\leq$~0.0090\\
He~I~$\lambda$5875		&	$\leq$~0.00263&	0.0773$\pm$0.0082&	0.0403$\pm$0.0027&	0.0618$\pm$0.0031&	0.0816$\pm$0.0043\\
$[$\ion{O}{1}] $\lambda$6300		&	$\leq$~0.00244&	0.270$\pm$0.026&	$\leq$~0.0057&	0.0283$\pm$0.0019&	0.1320$\pm$0.0055\\
$[$S\,\iii]~$\lambda$6312		&	$\leq$~0.0121&	$\leq$~0.0064&	$\leq$~0.0058&	$\leq$~0.0044&	$\leq$~0.0070\\
$[$\ion{O}{1}] $\lambda$6363		&	$\leq$~0.00232&	0.0707$\pm$0.0072&	0.0073$\pm$0.0018&	$\leq$~0.0042&	0.0294$\pm$0.0025\\
$[$N\,\ii]~$\lambda$6548		&	0.350$\pm$0.083&	1.45$\pm$0.10&	0.1760$\pm$0.0067&	0.310$\pm$0.012&	0.510$\pm$0.019\\
H$\alpha$ $\lambda$6562	&	2.63$\pm$0.60&	2.43$\pm$0.20&	3.09$\pm$0.10&	3.33$\pm$0.10&	3.09$\pm$0.10\\
$[$N\,\ii]~$\lambda$6583		&	1.15$\pm$0.30&	4.53$\pm$0.40&	0.630$\pm$0.023&	0.920$\pm$0.037&	1.528$\pm$0.057\\
He~I~$\lambda$6678		&	$\leq$~0.0099&	0.0285$\pm$0.0034&	0.0141$\pm$0.0017&	0.0253$\pm$0.0016&	0.0191$\pm$0.0021\\
$[$S\,\ii]~$\lambda$6716		&	0.250$\pm$0.060&	0.960$\pm$0.092&	0.1550$\pm$0.0059&	0.1550$\pm$0.0062&	0.430$\pm$0.016\\
$[$S\,\ii]~$\lambda$6730		&	0.200$\pm$0.048&	0.740$\pm$0.070&	0.1210$\pm$0.0047&	0.1290$\pm$0.0052&	0.360$\pm$0.013\\
He~I~$\lambda$7065		&	$\leq$~0.0034&	0.0089$\pm$0.0021&	$\leq$~0.00241&	0.00448$\pm$0.00064&	$\leq$~0.0056\\
$[$Ar~\iii]~$\lambda$7135		&	$\leq$~0.0036&	0.0369$\pm$0.0039&	$\leq$~0.00102&	0.00720$\pm$0.00061&	$\leq$~0.0057\\
$[$Fe~\iii]~$\lambda$7155		&	$\leq$~0.00080&	$\leq$~0.0053&	$\leq$~0.00113&	$\leq$~0.00040&	$\leq$~0.00218\\
He~I~$\lambda$7281		&	$\leq$~0.00080&	$\leq$~0.00198&	$\leq$~0.00102&	$\leq$~0.00040&	$\leq$~0.00217\\
$[$Ca~\ii]~$\lambda$7291	&	$\leq$~0.00080&	$\leq$~0.00198&	$\leq$~0.00102&	$\leq$~0.00040&	$\leq$~0.0056\\
$[$O~\ii]~$\lambda$7319		&	$\leq$~0.00080&	0.0300$\pm$0.0033&	$\leq$~0.00113&	$\leq$~0.0039&	0.0185$\pm$0.0059\\
$[$O~\ii]~$\lambda$7330		&	$\leq$~0.00080&	0.0197$\pm$0.0026&	0.00629$\pm$0.00086&	0.00752$\pm$0.00087&	$\leq$~0.0175\\
$[$N\,\iii]~$\lambda$7378	&	$\leq$~0.0032&	0.0168$\pm$0.0024&	$\leq$~0.00250&	$\leq$~0.00154&	0.0139$\pm$0.0019\\
$[$Ar~\iii]~$\lambda$7751		&	$\leq$~0.0078&	0.0170$\pm$0.0024&	0.0049$\pm$0.0016&	0.01000$\pm$0.00087&	0.0112$\pm$0.0011\\
P23 $\lambda$8392 	&	$\leq$~0.034&	$\leq$~0.0084&	$\leq$~0.0177&	0.0197$\pm$0.0023&	0.0367$\pm$0.0035\\
P22 $\lambda$346		&	$\leq$~0.033&	$\leq$~0.0186&	$\leq$~0.0095&	$\leq$~0.0066&	$\leq$~0.0105\\
P21 $\lambda$8359		&	$\leq$~0.036&	$\leq$~0.0199&	$\leq$~0.0096&	$\leq$~0.0060&	$\leq$~0.0105\\
P20 $\lambda$8374		&	$\leq$~0.036&	$\leq$~0.0196&	$\leq$~0.0177&	0.0077$\pm$0.0022&	0.0107$\pm$0.0035\\
P19 $\lambda$8413		&	$\leq$~0.034&	$\leq$~0.0090&	$\leq$~0.0178&	0.0103$\pm$0.0021&	0.0169$\pm$0.0034\\
P18 $\lambda$8437		&	$\leq$~0.032&	$\leq$~0.0090&	$\leq$~0.0093&	$\leq$~0.00199&	$\leq$~0.0047\\
O~I~$\lambda$8446		&	$\leq$~0.034&	$\leq$~0.0201&	$\leq$~0.0094&	0.0115$\pm$0.0023&	0.0158$\pm$0.0033\\
P17 $\lambda$8467		&	$\leq$~0.034&	$\leq$~0.0087&	$\leq$~0.0099&	$\leq$~0.0060&	$\leq$~0.0096\\
P16 $\lambda$8502		&	$\leq$~0.035&	$\leq$~0.0203&	$\leq$~0.0182&	$\leq$~0.00206&	$\leq$~0.0099\\
P15 $\lambda$8545		&	$\leq$~0.036&	$\leq$~0.0206&	$\leq$~0.0104&	$\leq$~0.0062&	$\leq$~0.0100\\
P14 $\lambda$8598		&	$\leq$~0.037&	$\leq$~0.0214&	$\leq$~0.0191&	$\leq$~0.0069&	$\leq$~0.0102\\
P13 $\lambda$8665		&	$\leq$~0.037&	$\leq$~0.0203&	$\leq$~0.0192&	0.0133$\pm$0.0022&	0.0132$\pm$0.0038\\
N~I~$\lambda$8683		&	$\leq$~0.037&	$\leq$~0.0218&	$\leq$~0.0193&	$\leq$~0.0069&	$\leq$~0.0113\\
P12 $\lambda$8750		&	$\leq$~0.037&	0.0321$\pm$0.0077&	$\leq$~0.0193&	0.0091$\pm$0.0023&	$\leq$~0.0114\\
P11 $\lambda$8862		&	$\leq$~0.036&	0.0314$\pm$0.0076&	$\leq$~0.0188&	0.0202$\pm$0.0025&	0.0246$\pm$0.0040\\
P10 $\lambda$9015		&	$\leq$~0.0205&	0.0163$\pm$0.0042&	0.0137$\pm$0.0032&	0.0187$\pm$0.0019&	0.0175$\pm$0.0035\\
$[$S\,\iii]~$\lambda$9068		&	$\leq$~0.0223&	0.0768$\pm$0.0081&	0.0216$\pm$0.0034&	0.1140$\pm$0.0045&	0.0653$\pm$0.0044\\
P9 $\lambda$9229		&	$\leq$~0.0201&	$\leq$~0.0123&	0.0222$\pm$0.0038&	0.0181$\pm$0.0020&	0.0128$\pm$0.0042\\
$[$S\,\iii]~$\lambda$9530		&	$\leq$~0.0131&	0.097$\pm$0.013&	0.0442$\pm$0.0067&	0.1430$\pm$0.0066&	0.0687$\pm$0.0082\\
P8 $\lambda$9546		&	$\leq$~0.040&	$\leq$~0.035&	0.0359$\pm$0.0075&	0.0493$\pm$0.0061&	0.044$\pm$0.013\\
C$_{H\beta}$		&	0.480$\pm$0.017&	0.390$\pm$0.013&	0.3860$\pm$0.0087&	0.6910$\pm$0.0096&	0.4530$\pm$0.0091\\
F$_{H\beta}$		&	57.09$\pm$0.80&	205.4$\pm$2.5&	170.0$\pm$1.5&	156.1$\pm$1.4&	125.9$\pm$1.1\\
EW$_{H\beta}$		&	16.5397	&	10.7871	&	21.9789 &	14.2089	&	12.5176\\
EW$_{H\alpha}$		&	90.4555	&	43.2975	&	204.974	&	106.304	&	62.5025\
\enddata
	\label{t:lineflux}
	\tablecomments{Full Table online only.}
\end{deluxetable*}

We report reddening-corrected line intensities measured from H\ii\ regions in the target fields in Table \ref{t:lineflux}.  The uncertainty associated with each measurement is determined from measurements of the spectral variance, extracted from the two-dimensional variance image, uncertainty associated with the flux calibration, Poisson noise in the continuum, read noise, sky noise, flat fielding calibration error, error in continuum placement, and error in the determination of the reddening.   We note that at the resolution of MODS, the [O\,\ii] $\lambda\lambda$3726; 3729 doublet is blended for all observations, however, the profile is clearly non-Gaussian in the majority of spectra.  We have modeled this doublet using two Gaussian profiles, for use primarily as a sanity check of the [S II] density determination.  For all calculations aside from this density check, we sum the flux in the [O\,\ii] $\lambda\lambda$3726; 3729 doublet. We note that we include a 2\% uncertainty based on the precision of the adopted flux calibration standards \citep[][see discussion in Berg et al. 2014]{Bohlin2010}.

\subsection{Diagnostic Diagrams}
\begin{figure}[bp] 
\epsscale{1.25}
   \centering
   \plotone{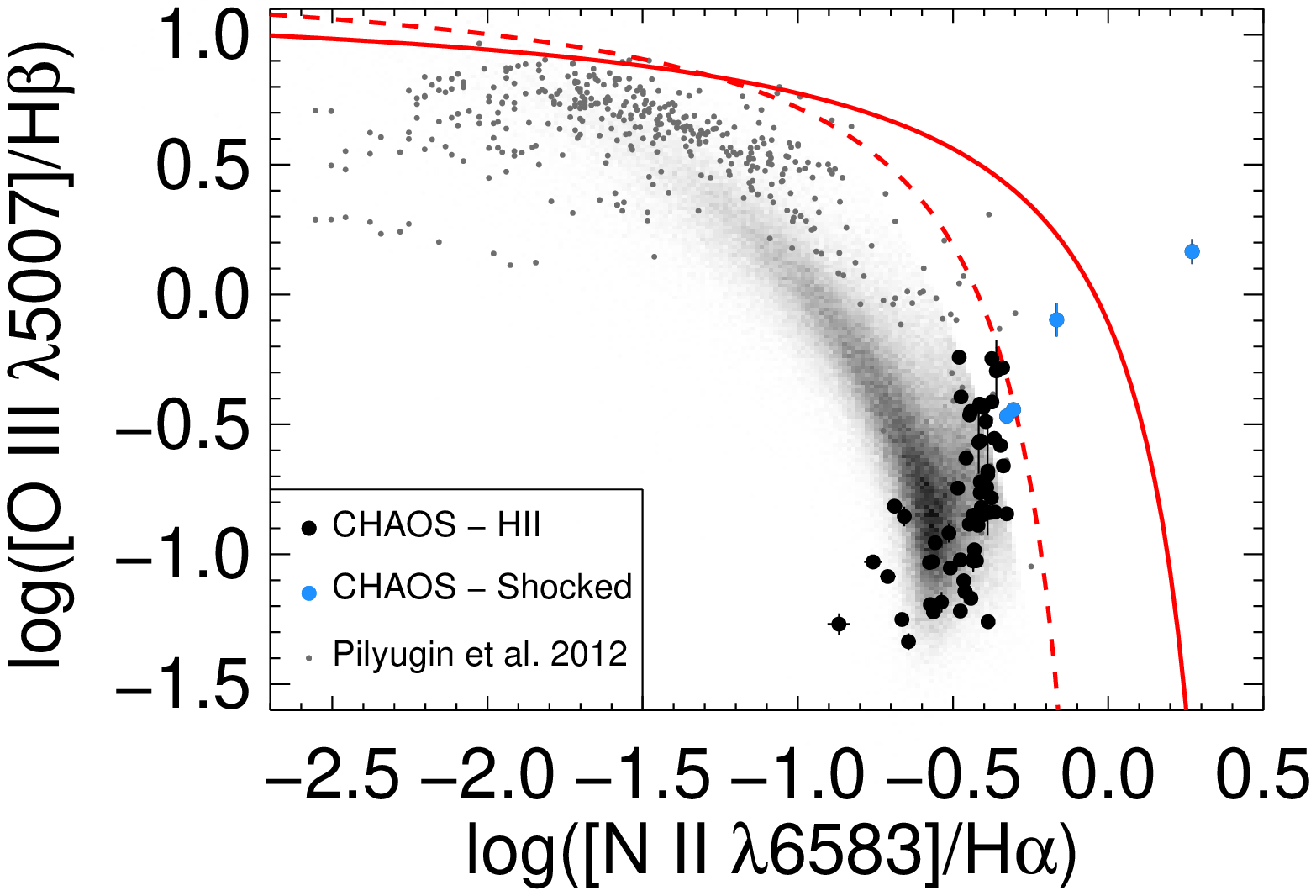}
   \plotone{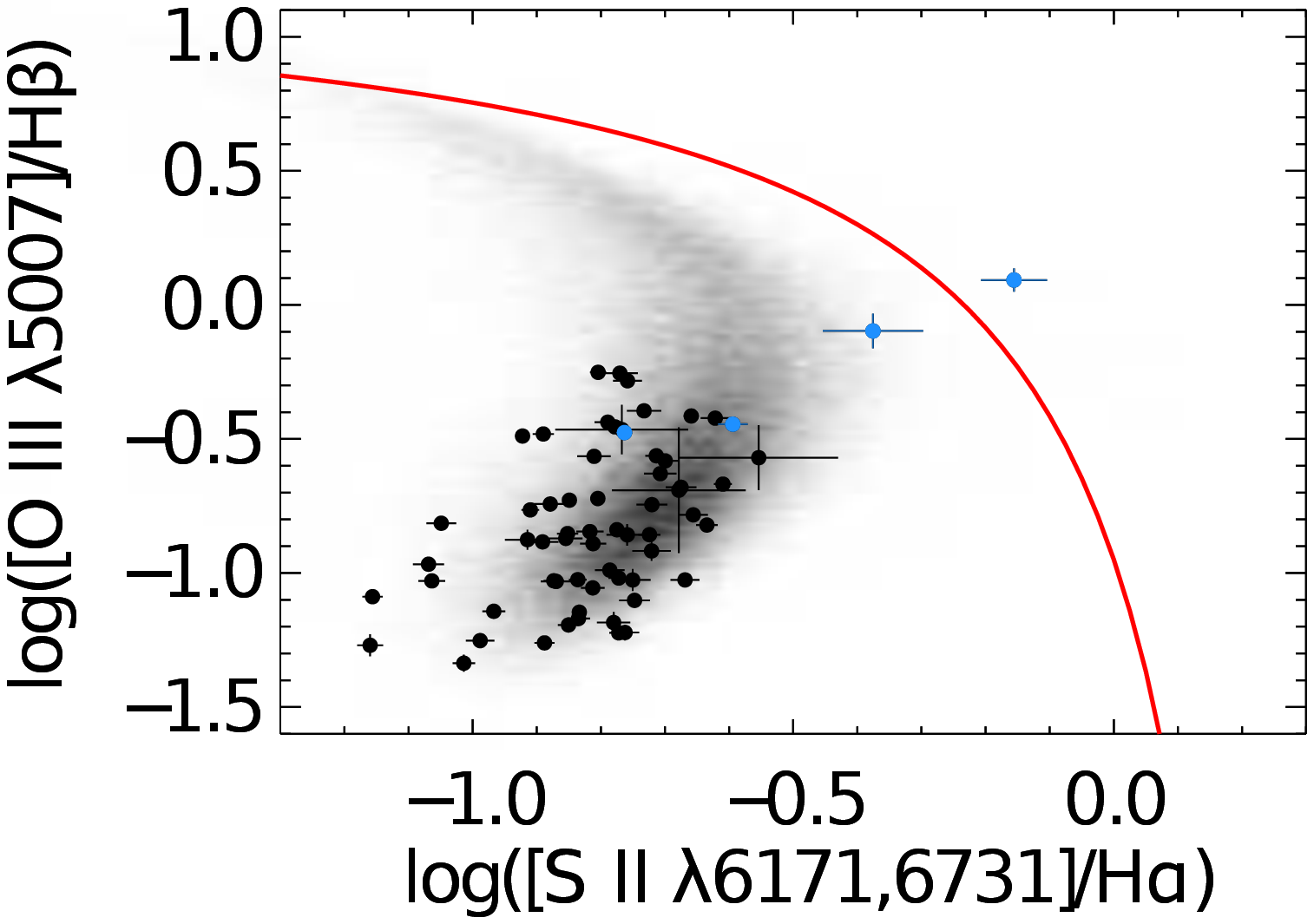}
   \plotone{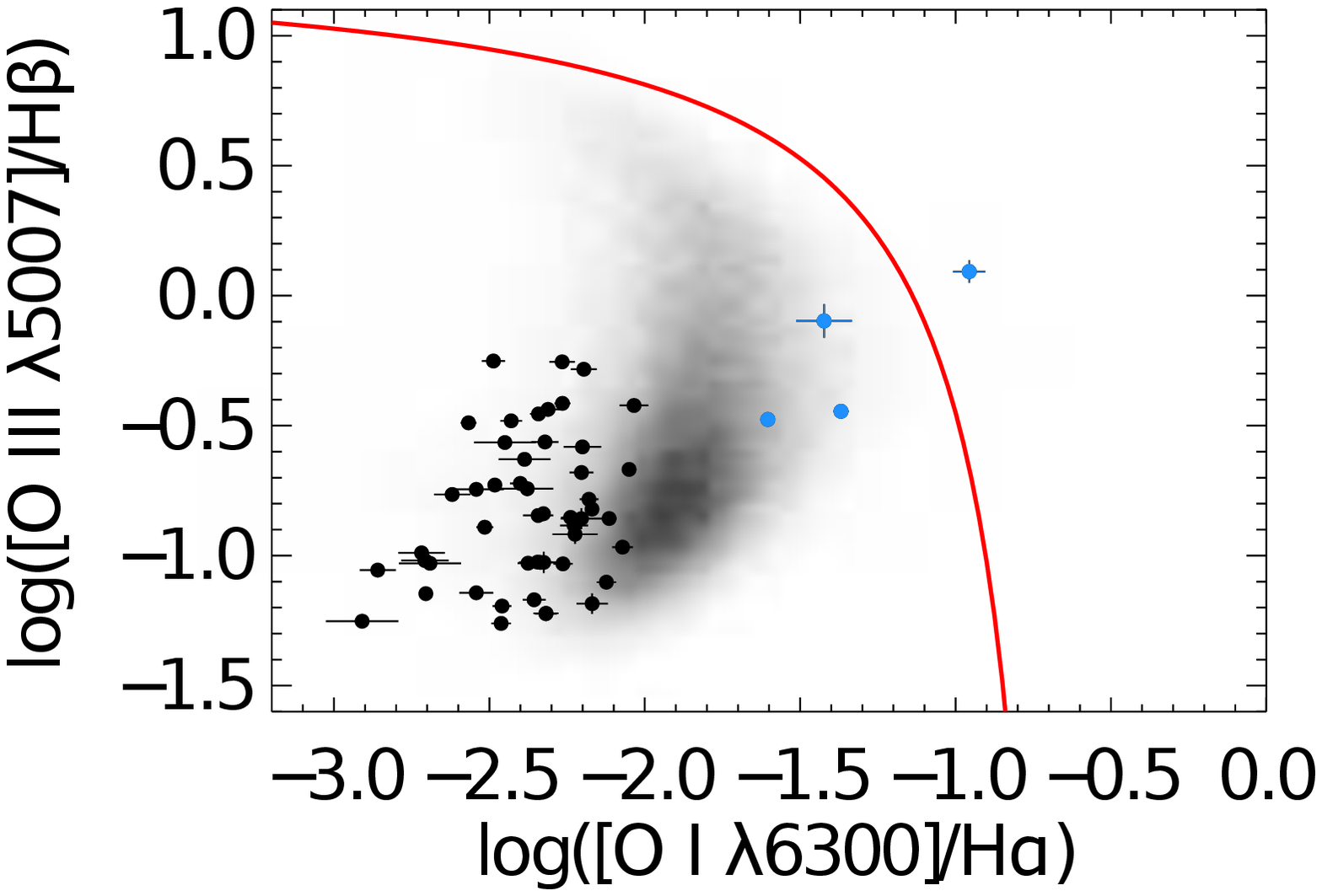}
   \caption{Diagnostic BPT plots composed of star forming SDSS galaxies (density plot), H\ii\ regions from \citep{pilyugin2012} (grey
squares), and regions observed in NGC~5194.  The dashed and solid red lines demarcate the boundaries between star forming galaxies and AGN as derived by \citet{kauffmann2003} and \citet{kewley2006}.  Blue points denote H\ii\ regions that display significant signs of shock ionization, i.e., [\ion{O}{1}] $\lambda$6300 $>$ 1\%, and are not included in the abundance analysis. Uncertainties in the strong line ratios measured from H\ii\ regions in NGC~5194 are roughly the size of the plotted data points when not visible.}  
   \label{fig:bpt}
\end{figure}
  
We selected target H\ii\ regions using H$\alpha$ imaging, which traces ionized gas.  However, other objects besides H\ii\ regions can also produce H$\alpha$ emission.  Notably, \citet{maddox2007} used the Very Large Array to map and measure the spectral index in 107 compact radio sources in NGC~5194.  Several radio sources that exhibit a steep spectral index, which is typical of supernovae remnants, lie along the inner spiral arm where some of our targeted H\ii\ regions are found. Thus, to verify that we are measuring lines from photoionized  H\ii\ regions, we plot standard diagnostic diagrams in Figure \ref{fig:bpt} \citep[BPT,][]{baldwin1981}, [O \iii]/H$\beta$ against [N\,\ii]/H$\alpha$ and [O \iii]/H$\beta$ against [\ion{O}{1}]/H$\alpha$.  H\ii\ regions from the current study are shown as black points while the density plot and small grey points indicate the locations of star forming galaxies presented in \citet{pilyugin2012} and star forming SDSS
galaxies\footnote{Available at http://www.mpa-garching.mpg.de/SDSS/DR7/} from the MPA-JHU catalog, respectively, in this diagnostic space. While most of the regions we observed do indeed exhibit the expected properties of both photoionization and the general locus of metal rich galaxies in these diagnostic diagrams, line ratios observed in four regions indicate possible signatures of shock ionization. 

Four regions, +2.5+9.5, $-$6.9+20.8, +30.2+2.2, and +13.3$-$141.3, have strong very strong [\ion{O}{1}] emission that is indicative of the presence of shock excitation causing these regions to deviate from the locus of star forming galaxies.  We detect temperature-sensitive auroral lines in two of these regions.  This includes our only detections of the [O~\iii] $\lambda$4636 line, which is normally detected in hotter, metal-poor regions.  Additionally, we detect Wolf-Rayet features, which are preferentially seen in metal rich regions \citep{massey2003,bonanos2010}, in one of these regions.  These features indicate the presence of a hot component, despite the dominant appearance of low-ionization features in these nebula (e.g., very low [O~\iii] $\lambda$5007/H$\beta$).  In two of these regions, in addition to strong [\ion{O}{1}] emission we also detect strong [S\,\ii] and [N\,\ii] emission which is characteristic of supernovae remnants \citep{skillman1985}.  Given the possible presence of shocks, whether from supernovae or Wolf-Rayet star winds, these regions cannot be interpreted as purely photoionized H\ii\ regions \citep{binette2012}.  We note that \citet{bresolin2002} found that the presence of evolved massive stars does not significantly affect the ionizing radiation field of embedded star clusters.  Given that most regions in which we detect Wolf-Rayet features (11 out of 12), do not show signs of shock excitation, the shocks may have an alternate origin, but still indicate extra sources of ionization.  We have therefore excluded these four regions from the primary abundance analysis; however, given the relative insensitivity of the N/O ratio to temperature, we include these regions in our analysis of N/O.  We report the derived temperatures and abundances for the two shock regions with auroral line detections in Table \ref{t:shocktable}.

\begin{deluxetable}{cccccccc}  
	\tabletypesize{\scriptsize}
	\tablecaption{N/O in Shocked Regions in NGC 5194}
	\tablewidth{0pt}
	\tablehead{   
	  \colhead{}	&
	  \colhead{NGC5194+2.5+9.5}	&
	  \colhead{NGC5194+30.2+2.2}	
	  }
\startdata
T[O\,\ii]\,measured (K) 	&	6880$\pm$340&	8740$\pm$150\\
T[O\,\iii]\,measured (K) 	&	19000$\pm$2100&	22100$\pm$2000\\
T[N\,\ii]\,measured (K) 	&	10300$\pm$550&	9100$\pm$210\\
T[S\,\iii]\,measured (K) 	&	\ldots&	7550$\pm$660\\
\vspace{.1cm}			\
n($_e$)\,measured (cm$^{-3}$) 	&	110$\pm$130&	362$\pm$44\\
T[O\,\ii]used (K) 	&	10300$\pm$550&	9100$\pm$210\\
\vspace{.1cm}			\
n($_e$)\,adopted (cm$^{-3}$) 	&	109.$\pm$132.&	362.$\pm$44.\\
O+/H+ (10$^{5}$)	&	19.3$\pm$2.9&	9.36$\pm$0.50\\
N+/H+ (10$^{6}$)	&	77.2$\pm$7.3&	33.3$\pm$1.1\\
log(N/O) (dex)	&	-0.400$\pm$0.077&	-0.450$\pm$0.029\\
\enddata
	\label{t:shocktable}
	\tablecomments{Regions showing signs of shock ionization.\\}
\end{deluxetable}

\section{Gas-Phase Abundances}
Elemental abundances can be determined from emission line spectra given (1) the electron density (n$_e$), (2) the electron temperature (T$_e$), and (3) a correction factor for unobserved ionic states.  When auroral lines are detected, we follow the methodology of \citet{agn3}, utilizing the fact that these lines, paired with their stronger ionic counterparts, are very sensitive to electron temperature. If T$_e$ cannot be \emph{directly} calculated then abundances must be derived using less precise indirect methods, wherein measurements of the strong-lines have been calibrated either empirically or via photoionization modeling \citep[e.g.,][]{edmunds1984}.   In calculating temperatures and abundances we have adopted the atomic data presented \citet{berg2014}.  \\

\subsection{Direct Abundances}
We use the [S\,\ii] $\lambda$6717/6731 line ratio to determine n$_e$ in each region as these lines are both relatively strong and well separated.  Most H\ii\ regions in this study, 43 out of 59,  have an [S\,\ii] ratio that lies within the low-density regime [I($\lambda$6717)/I($\lambda$6731) $>$ 1.35].  For these regions we adopt an n$_e$ of 100 cm$^{-3}$.  For the remaining 16 regions we calculate electron densities using a five level atom code based on FIVEL \citep{fivel}, see Table \ref{t:labundances}.  Although these regions have a [S\,\ii] ratio that falls below the selected limit, they are still lie in the low-density regime as the highest density determined is 227 cm$^{-3}$.

\begin{deluxetable*}{cccccccc}  
	\tabletypesize{\scriptsize}
	\tablecaption{Abundances in NGC 5194}
	\tablewidth{0pt}
	\tablehead{   
	  \colhead{}	&
	  \colhead{NGC5194-4.3+63.3 }	&
	  \colhead{NGC5194-33.2+58.0 }	&
	  \colhead{NGC5194-62.2+50.3 }	&
	  \colhead{NGC5194+75.5-28.7 }	&
	  \colhead{NGC5194-98.9-52.8  }	
	  }
\startdata
T[O\,\ii]measured (K)  	&	8360$\pm$710&	7330$\pm$480&	8020$\pm$390&	7890$\pm$520&	6780$\pm$660\\
T[O\,\iii]measured (K)  &		\ldots&	\ldots&	\ldots&	\ldots&	\ldots\\
T[N\,\ii]measured (K)  	&	5700$\pm$320&	5740$\pm$390&	6380$\pm$430\ldots&	5980$\pm$260\\
T[S\,\iii]measured (K) &		\ldots&	\ldots&	\ldots&	5630$\pm$560&	\ldots\\
\vspace{.1cm}							
n($_e$)measured (cm$^{-3}$) 	&	124$\pm$41&	153$\pm$44&	78$\pm$33&	106$\pm$40&	43$\pm$34\\
T[O\,\ii]used (K) 	&	5700$\pm$320&	5740$\pm$390&	6380$\pm$430&	6320$\pm$450&	5980$\pm$260\\
T[O\,\iii]used (K) 	&	3810$\pm$470&	3870$\pm$580&	4810$\pm$640&	4740$\pm$670&	4230$\pm$380\\
T[N\,\ii]used (K) 	&	5700$\pm$320&	5740$\pm$390&	6380$\pm$430&	6320$\pm$450&	5980$\pm$260\\
T[S\,\iii]used (K) 	&	4870$\pm$390&	4910$\pm$480&	5700$\pm$530&	5630$\pm$560&	5210$\pm$320\\
\vspace{.1cm}							
n($_e$)adopted (cm$^{-3}$) 	&	124.$\pm$41.&	153.$\pm$44.&	100.$\pm$100.&	106.$\pm$40.&	100.$\pm$100.\\\hline
O+/H+ (10$^{5}$)	&	27.4$\pm$5.2&	36.7$\pm$8.4&	32.7$\pm$6.6&	28.7$\pm$6.2&	34.1$\pm$5.2\\
O++/H+ (10$^{5}$) 	&	41$\pm$20&	32$\pm$19&	20.3$\pm$8.4&	20.3$\pm$9.2&	21.5$\pm$6.9\\
\vspace{.1cm}							
12 + log(O/H) (dex)	&	8.84$\pm$0.10&	8.84$\pm$0.10&	8.724$\pm$0.088&	8.691$\pm$0.098&	8.745$\pm$0.067\\\hline
N+/H+ (10$^{6}$)	&	101$\pm$11&	155$\pm$21&	94$\pm$12&	95$\pm$12&	108.2$\pm$9.2\\
N ICF 	&	2.50$\pm$0.90&	1.88$\pm$0.70&	1.62$\pm$0.50&	1.71$\pm$0.50&	1.63$\pm$0.40\\
log(N/O) (dex)	&	-0.430$\pm$0.095&	-0.40$\pm$0.10&	-0.50$\pm$0.10&	-0.50$\pm$0.10&	-0.500$\pm$0.076\\
\vspace{.1cm}							
12 + log(N/H) (dex)	&	8.40$\pm$0.20&	8.46$\pm$0.20&	8.18$\pm$0.10&	8.21$\pm$0.10&	8.25$\pm$0.10\\\hline
S+/H+ (10$^{7}$) 	&	69.1$\pm$7.5&	64.0$\pm$8.3&	39.1$\pm$4.6&	46.9$\pm$5.9&	43.5$\pm$3.6\\
S++/H+ (10$^{7}$)	&	39.2$\pm$5.7&	115$\pm$19&	54.4$\pm$8.2&	79$\pm$12&	109$\pm$11\\
S ICF 	&	1.26$\pm$0.10&	1.19$\pm$0.10&	1.14$\pm$0.10&	1.16$\pm$0.10&	1.14$\pm$0.10\\
log(S/O) (dex)  	&	-1.70$\pm$0.10&	-1.51$\pm$0.10&	-1.70$\pm$0.10&	-1.53$\pm$0.10&	-1.504$\pm$0.087\\
\vspace{.1cm}							
12 + log(S/H) (dex)	&	7.133$\pm$0.058&	7.326$\pm$0.067&	7.027$\pm$0.062&	7.164$\pm$0.063&	7.241$\pm$0.054\\\hline
Ne++/H+ (10$^{6}$)	&	\ldots&	\ldots&	\ldots&	\ldots&	\ldots\\
Ne ICF	&	1.66$\pm$1.00&	2.1$\pm$1.4&	2.6$\pm$1.2&	2.4$\pm$1.2&	2.58$\pm$0.90\\
log(Ne/O) (dex)&		\ldots&	\ldots&	\ldots&	\ldots&	\ldots\\
\vspace{.1cm}							
12 + log(Ne/H) (dex) &		\ldots&	\ldots&	\ldots&	\ldots&	\ldots\\\hline
Ar++/H+ (10$^{7}$) 	&	4.9$\pm$1.3&	16.0$\pm$3.3&	19.0$\pm$3.4&	9.3$\pm$1.9&	14.0$\pm$2.0\\
Ar ICF 	&	1.46$\pm$0.10&	1.48$\pm$0.10&	1.62$\pm$0.20&	1.55$\pm$0.20&	1.61$\pm$0.20\\
log(Ar/O) (dex)	&	-2.98$\pm$0.20&	-2.46$\pm$0.20&	-2.24$\pm$0.10&	-2.53$\pm$0.10&	-2.39$\pm$0.10\\
12 + log(Ar/H) (dex) 	&	5.85$\pm$0.10&	6.37$\pm$0.10&	6.487$\pm$0.089&	6.159$\pm$0.098&	6.353$\pm$0.076
\enddata
	\label{t:labundances}
	\tablecomments{Full Table online only.\\}
\end{deluxetable*}

As NGC~5194 is generally metal rich, we do not detect the most common temperature-sensitive line, [O~\iii] $\lambda$4363.  While we can place upper limits on the flux of this line, given the strong dependence on temperature in this regime, those limits do not translate into useful limits on gas-phase abundances.  Rather, we focus on the [N\,\ii] $\lambda$5755 and [S\,\iii] $\lambda$6312 auroral lines.  We only use the most reliable spectra and require the auroral lines to have an amplitude greater than three times the rms noise measured in the local continuum.  We also ensured the presence of a robustly measurable line via visual inspection of the spectra.  From our sample of 59 observed H\ii\ regions, we detect [N\,\ii] $\lambda$5755 in 26 regions and [S\,\iii] $\lambda$6312 in 17 regions.  We note that 15 of the regions where we detect [S\,\iii] $\lambda$6312 we also detect  [N\,\ii] $\lambda$5755 (see Table \ref{t:locations}), for a total of 28 H\ii\ regions with [S\,\iii] or [N\,\ii] auroral line detections.

We represent the temperature structure of an H\ii\ region with a three-zone model wherein each zone is characterized by a different T$_e$. We adopt for the ionization stages represented in each zone: [O\,\ii], [N\,\ii], [S\,\ii] in the low-ionization zone; [S\,\iii], [Ar\,\iii] in the intermediate-ionization zone; and [O \iii], [Ne \iii] in the high-ionization zone.  When both [N\,\ii] $\lambda$5755 and [S\,\iii] $\lambda$6312 are detected in a single H\ii\ region we adopt the derived temperatures for their respective zones.  To calculate the temperature of the high ionization zone, or the temperature of either the intermediate or low zone when only a single auroral line is detected, we use the scaling relations of \citet{garnett1992}:
\begin{equation} \rm{T[N~II] = T[O~II] = T[S~II] = }0.70 \rm{T[O~III]} + 3000 \rm{\,K} \end{equation}
\begin{equation} \rm{T[S~III] = }0.83\rm{T[O~III]} + 1700 \rm{\,K.} \end{equation}
We note that the relation derived between T[N\,\ii] and T[O\,\iii] by \citet{pilyugin2009} is similar to the relation between these two temperatures derived by \citet{garnett1992}.  The selection of one relation or another does not significantly alter our results.  Measured temperatures and the adopted temperatures are reported in Table \ref{t:labundances}.  

\begin{figure}[tbp] 
\epsscale{1.28}
   \centering
   \plotone{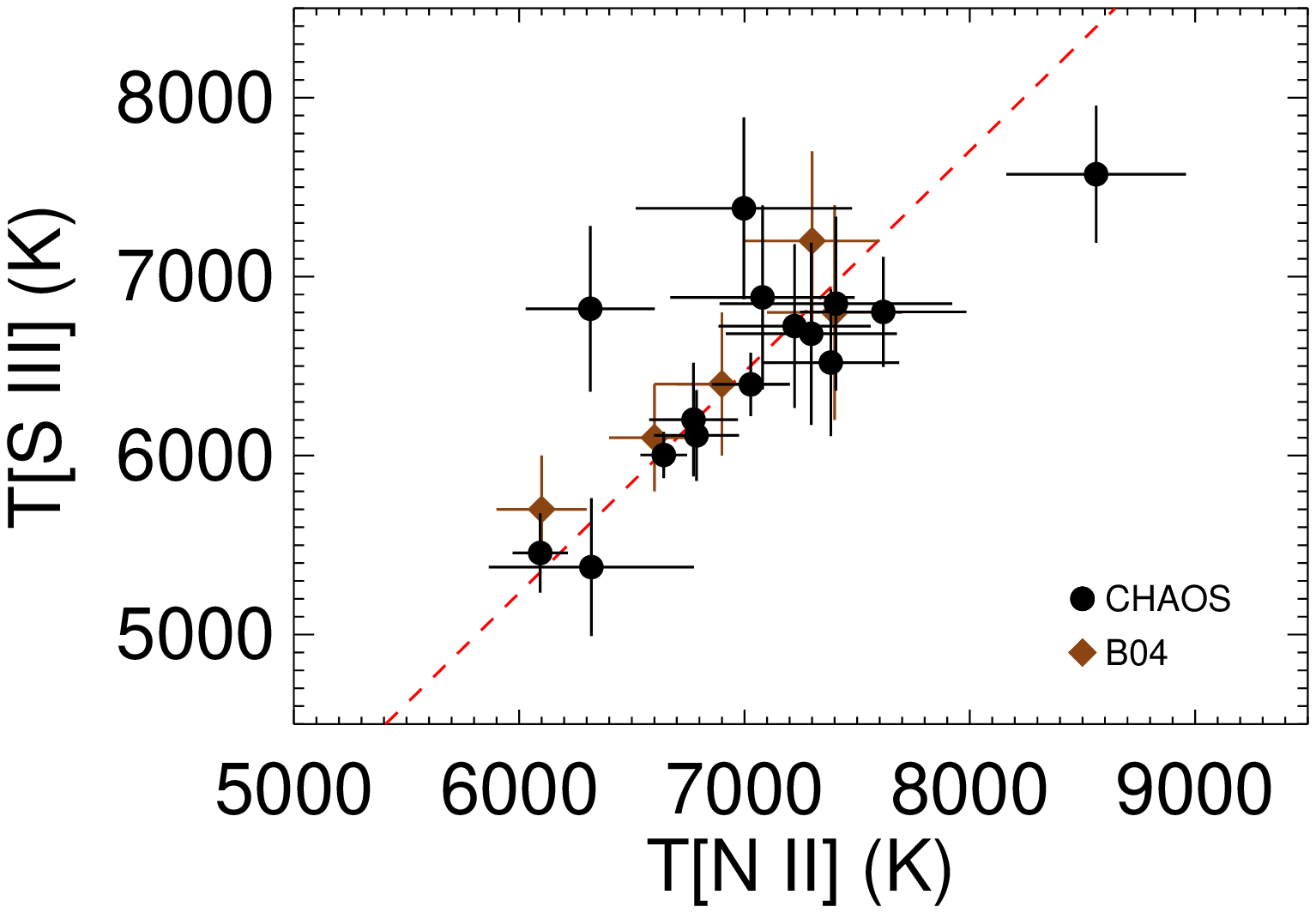}
   \plotone{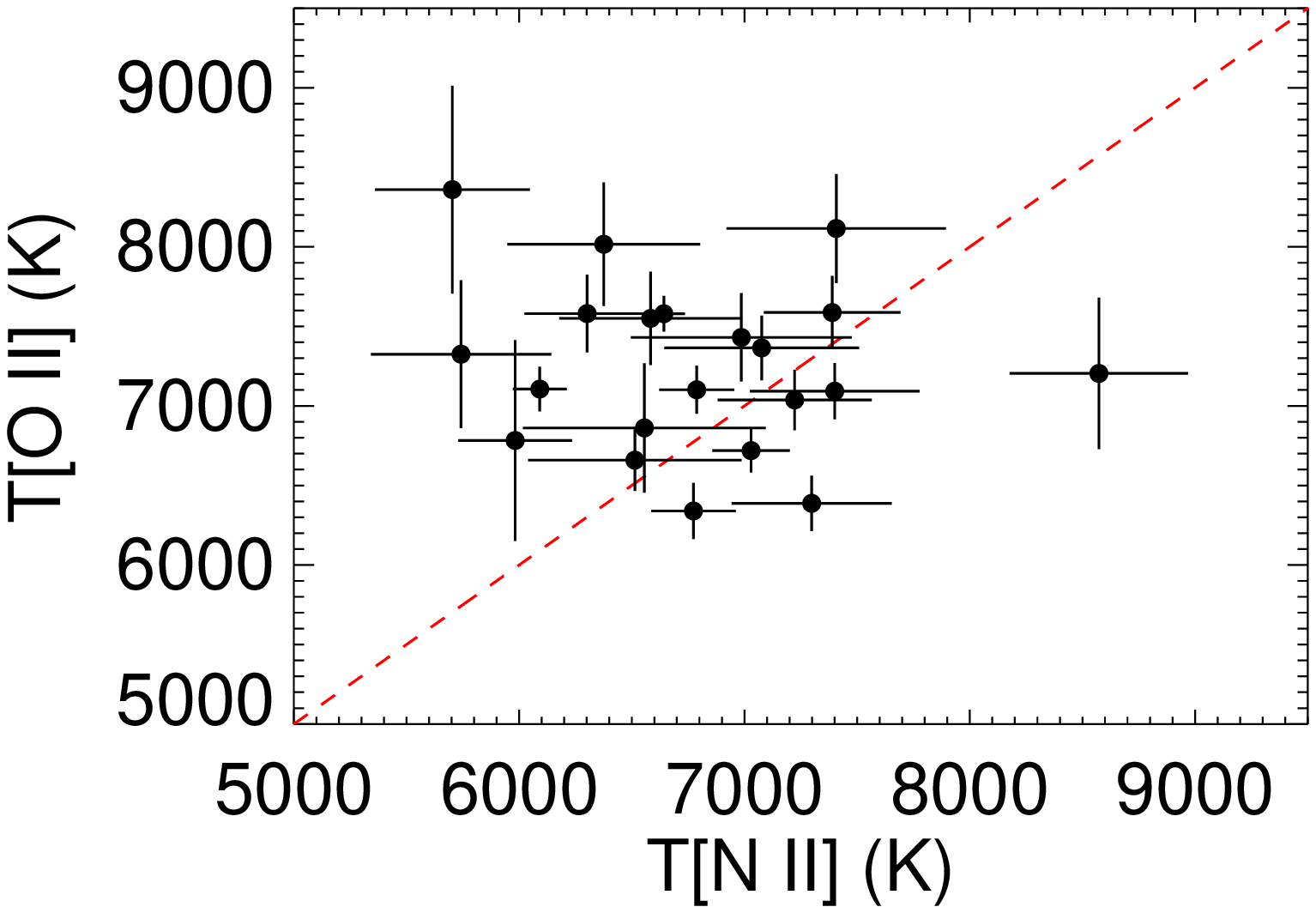}
   \plotone{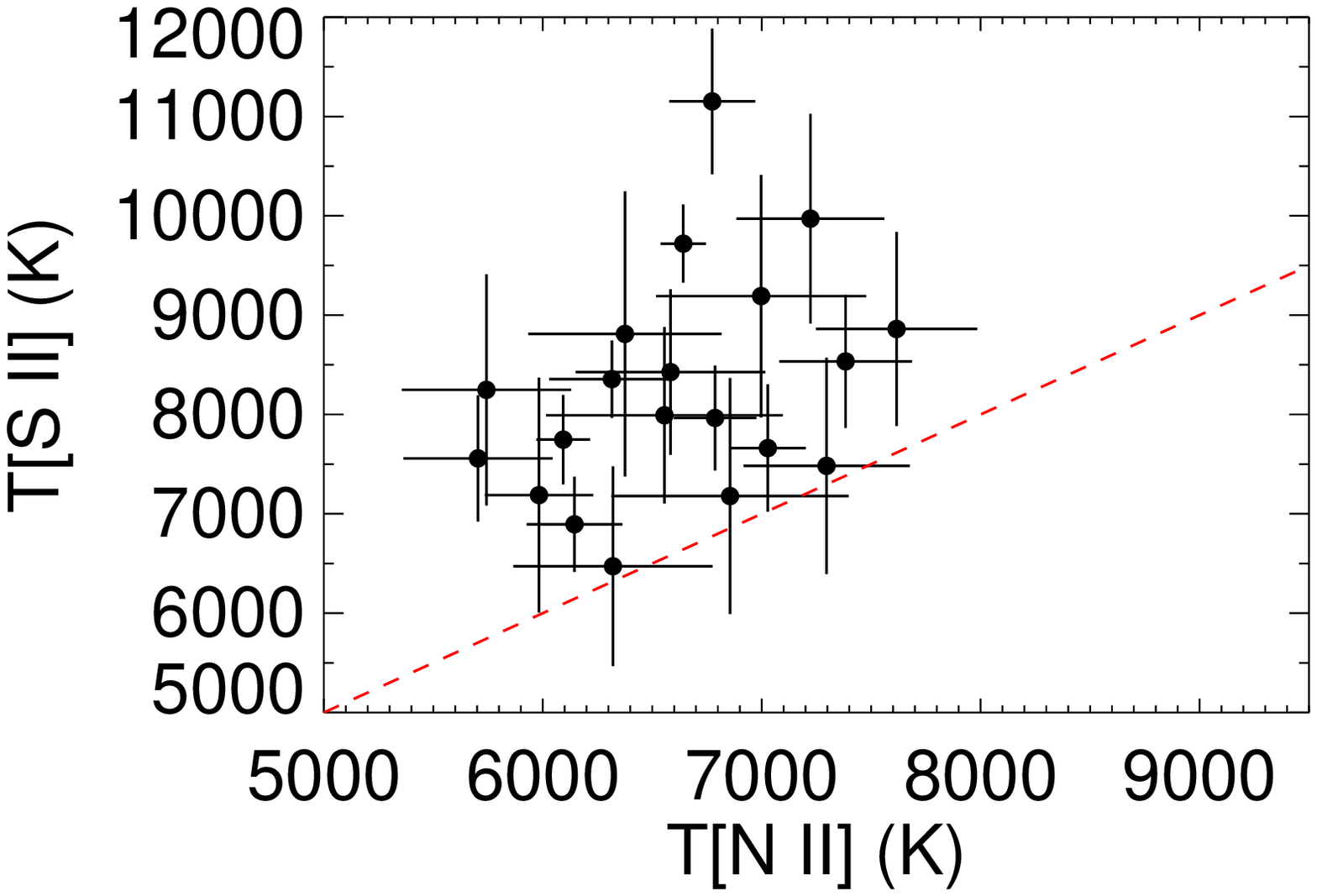}
   \caption{Relationship between T[N\,\ii] and T[S\,\iii] (Top), T[N\,\ii] and T[O~\ii] (middle), T[N\,\ii] and T[S\,\ii] (Bottom) measured for H\ii\ regions in NGC~5194.  The model predictions from \citet{garnett1992} are drawn as a dashed line.  Open brown diamonds in the top panel are from \citet{bresolin2004}}  
   \label{fig:temperature}
\end{figure}  

To check the validity of the adopted scaling relations, we show the relationship between T[N\,\ii] and T[S\,\iii]  in Figure \ref{fig:temperature} along with Equation 2 (red dashed line).  We find an excellent agreement between our T[S\,\iii] and T[N\,\ii] values with the scaling relationship of G92 \citep[as has been previously observed by][]{kennicutt2003}.  Additionally, \citet{bresolin2004} present independent T[N\,\ii] determinations for eight H\ii\ regions in common with our sample.  We find good agreement for all but one region with a mean offset of $\Delta$T$\sim-10$\,K, which is statistically negligible given the scatter.  In the discrepant region, +83.4$-$133.1 / CCM57, we measure T[N\,\ii] to be $\sim600$~K smaller than \citet{bresolin2004}.  Comparing our directly determined T[S\,\iii] with \citet{bresolin2004} yields an offset that is slightly larger than the offset seen in T[N\,\ii], $\sim-300$\,K, but still consistent within the errors.  Data from \citet{bresolin2004} for which both T[N\,\ii] and T[S\,\iii] are measured are also shown in Figure \ref{fig:temperature} as open brown diamonds.  

In addition to the [N\,\ii] $\lambda$5755 and [S\,\iii] $\lambda$6312 auroral lines, we detect [O~\ii] $\lambda$7319,7330 and [S\,\ii] $\lambda$4068,4076 in several regions (see Table \ref{t:locations}).  Given that these weak doublets both lie on top of stellar absorption features, their fitted line strengths are highly dependent on the model stellar continuum.  Furthermore, the [O~\ii] doublet is complicated by sky-subtraction as it lies just inside the blue-end of strong OH Meinel band emission.  Nevertheless, we report temperatures determined from these lines in Table \ref{t:labundances}.  In general, temperatures derived from the [O\,\ii] lines are consistent with the cool temperatures found using both [N\,\ii] $\lambda$5755 and [S\,\iii] $\lambda$6312, but do show large scatter, similar to the findings of \citet{kennicutt2003}.  On the other hand, the temperatures we derive using the [S\,\ii] lines are biased to higher temperatures compared to both [N\,\ii] $\lambda$5755 and [S\,\iii] $\lambda$6312 and show significant scatter.  Thus, we adopt temperatures derived from [N\,\ii] $\lambda$5755 and [S\,\iii] $\lambda$6312 except in the one H\ii\ region where the [O\,\ii] doublet was the only temperature-sensitive feature detected. 

H\ii\ regions where we do not robustly detect an auroral line are spread throughout the disk of NGC~5194.  There is, however, a clustering of H\ii\ regions at radii of 0.06 $<$ R/R$_{25}$ $<$ 0.11 that that all lie on the north-side of the nucleus of NGC~5194.  These regions lie interior to all regions where we measured auroral lines, excluding two objects with unusually strong [\ion{O}{1}] emission.  We combined the one-dimensional spectra from these nine regions to create an average spectrum of these inner H\ii\ regions.  Notwithstanding the high signal-to-noise combined spectrum\footnote{This is equivalent to an 18 hour exposure on an 8.3m telescope.}, we only marginally detect [N\,\ii] $\lambda$5755 and do not detect [S\,\iii] $\lambda$6312.  This measurement is complicated by the presence of the stellar bulge of NGC~5194.  This old stellar population exhibits an absorption feature underneath [N\,\ii] $\lambda$5755.  Thus, a direct abundance from this line is strongly coupled to the adopted stellar continuum model.  Furthermore, as these H\ii\ regions are quite metal rich, they are consistent with a very cool temperature, T[N~II] $\approx$ 5700~K, where the temperature-sensitive [N\,\ii] ratio, I($\lambda$5755)/[I($\lambda$6548) + I($\lambda$6583)], is sensitive to small changes in temperature.  While our direct abundance determination from this composite spectrum is less certain due to the complications in measuring [N\,\ii] $\lambda$5755, we include it in our analysis as it provides a constraint on the oxygen abundance in the inner portion of the galaxy. 

While emission lines are measured for all dominant ionization states of oxygen, derivation of abundances for other elements requires us to account for the presence of unobserved ionization states. Nitrogen abundances were derived under the assumption N/O = N$^+$/O$^+$; neon abundances were derived under the assumption Ne/O = Ne$^{++}$/O$^{++}$ \citep{peimbert1969}. In the cases of both sulfur and argon, we adopt the analytical ionization correction factors of \citet{thuan1995}, based on the photoionization modeling of \citet{stasinska1990}.  We report all ionization correction factors and abundance ratios in Table \ref{t:labundances}.    

\subsection{Strong-Line Abundances}
While the absolute calibration of strong-line abundances are uncertain \citep{kewley2008}, relative abundances which depend on ionic ratios which are less sensitive to T$_e$ can still be robustly determined in H\ii\ regions where the auroral lines are not detected.  While a full analysis of all strong line methods will be deferred to a future paper with the full CHAOS dataset\footnote{We prefer to undertake this with the a more complete sample from the CHAOS project so that we can thoroughly cover the full abundance range, rather than only the metal rich end.}, we employ a semi-empirical approach \citep{vanzee1997} where an electron temperature consistent with an adopted strong-line oxygen abundance is used to calculate the relative N/O abundance ratio.  We emphasize that this ratio is insensitive to changes in temperature because the temperature sensitivities between N$^+$ and O$^+$ are very similar.  We report the N/O values, and the assumed temperature and density in Table \ref{t:strongline_no}.

\begin{deluxetable*}{lcccc}  
	\tabletypesize{\scriptsize}
	\tablecaption{N/O in NGC 5194}
	\tablewidth{0pt}
	\tablehead{   
	  \colhead{}	&
	  \colhead{n($_e$)measured (cm$^{-3}$)}	&
	  \colhead{n($_e$)adopted (cm$^{-3}$)}	&
	  \colhead{T[N\,\ii]adopted (K)}	&
	  \colhead{log(N/O) (dex)}	
	  }
\startdata
NGC5194+1.1+19.8 	&	123$\pm$59&	123.$\pm$59.&	6200$\pm$1000&	-0.50$\pm$0.20	\\
NGC5194+0.2+20.2 	&	212$\pm$70&	212.$\pm$70.&	6000$\pm$1000&	-0.30$\pm$0.20	\\
NGC5194-6.9+20.8 	&	213$\pm$67&	213.$\pm$67.&	7000$\pm$1000&	-0.50$\pm$0.20	\\
NGC5194-18.5+13.7 	&	101$\pm$53&	101.$\pm$53.&	6000$\pm$1000&	-0.50$\pm$0.20	\\
NGC5194+19.7+15.0 	&	119$\pm$51&	119.$\pm$51.&	5900$\pm$1000&	-0.50$\pm$0.20	\\
NGC5194+27.2+3.6 	&	150$\pm$41&	150.$\pm$41.&	5900$\pm$1000&	-0.30$\pm$0.20	\\
NGC5194+21.3+19.6 	&	89$\pm$46&	100.$\pm$100.&	5600$\pm$1000&	-0.20$\pm$0.20	\\
NGC5194-27.9-18.1 	&	132$\pm$44&	132.$\pm$44.&	5900$\pm$1000&	-0.40$\pm$0.20	\\
NGC5194+16.3+30.8 	&	157$\pm$60&	157.$\pm$60.&	5800$\pm$1000&	-0.50$\pm$0.20	\\
NGC5194+17.7+30.2 	&	110$\pm$44&	110.$\pm$44.&	5600$\pm$1000&	-0.40$\pm$0.20	\\
NGC5194+4.1+56.5 	&	91$\pm$56&	100.$\pm$100.&	5600$\pm$1000&	-0.30$\pm$0.20	\\
NGC5194+42.7-55.8 	&	68$\pm$47&	100.$\pm$100.&	6400$\pm$1000&	-0.60$\pm$0.20	\\
NGC5194+62.7-36.9 	&	7.0$\pm$50 &	100.$\pm$100.&	6200$\pm$1000&	-0.40$\pm$0.20	\\
NGC5194-65.0+30.5 	&	49$\pm$50&	100.$\pm$100.&	6400$\pm$1000&	-0.70$\pm$0.20	\\
NGC5194+81.9-5.4 	&	99$\pm$52&	100.$\pm$100.&	6000$\pm$1000&	-0.50$\pm$0.20	\\
NGC5194+93.6+5.9 	&	83$\pm$56&	100.$\pm$100.&	6200$\pm$1000&	-0.50$\pm$0.20	\\
NGC5194-96.5-4.5 	&	46$\pm$27&	100.$\pm$100.&	6100$\pm$1000&	-0.60$\pm$0.20	\\
NGC5194+96.1+16.8 	&	33$\pm$38&	100.$\pm$100.&	5900$\pm$1000&	-0.60$\pm$0.20	\\
NGC5194+83.6+82.0 	&	43$\pm$42&	100.$\pm$100.&	5800$\pm$1000&	-0.50$\pm$0.20	\\
NGC5194-110.3-31.3 	&	37$\pm$28&	100.$\pm$100.&	5900$\pm$1000&	-0.60$\pm$0.20	\\
NGC5194-59.0-121.4 	&	53$\pm$33&	100.$\pm$100.&	6200$\pm$1000&	-0.60$\pm$0.20	\\
NGC5194-60.0-121.3 	&	56$\pm$45&	100.$\pm$100.&	6400$\pm$1000&	-0.70$\pm$0.20	\\
NGC5194+77.5+108.4 	&	43$\pm$44&	100.$\pm$100.&	6700$\pm$1000&	-0.60$\pm$0.20	\\
NGC5194+12.0-140.7 	&	33$\pm$48&	100.$\pm$100.&	6800$\pm$1000&	-0.70$\pm$0.20	\\
NGC5194+13.3-141.3 	&	20$\pm$100&	100.$\pm$100.&	8100$\pm$1000&	-0.60$\pm$0.20	\\
NGC5194-129.7+28.3 	&	29$\pm$57&	100.$\pm$100.&	6900$\pm$1000&	-0.80$\pm$0.20	\\
NGC5194+65.0+127.7 	&	3.0$\pm$30 &	100.$\pm$100.&	6500$\pm$1000&	-0.60$\pm$0.20	\\
NGC5194+92.3-118.5 	&	29$\pm$45&	100.$\pm$100.&	6300$\pm$1000&	-0.70$\pm$0.20	\\
NGC5194+146.4-6.7 	&	31$\pm$38&	100.$\pm$100.&	6800$\pm$1000&	-0.70$\pm$0.20	\\
NGC5194+147.7-7.6 	&	34$\pm$44&	100.$\pm$100.&	7200$\pm$1000&	-0.70$\pm$0.20	\\
NGC5194+63.1-149.7 	&	51$\pm$48&	100.$\pm$100.&	7000$\pm$1000&	-0.70$\pm$0.20	
\enddata
	\label{t:strongline_no}
	\tablecomments{N/O abundances for regions where no auroral lines were measured.  This ratio is insensitive to the temperature adopted, nevertheless we report the adopted semi-empirical temperature.}
\end{deluxetable*}

\subsection{Comparison with Previous Work}
Previous direct abundances were determined for NGC~5194 by \citet{bresolin2004}, who detected [N\,\ii] $\lambda$5755 in 10 regions, eight of which were included in the CHAOS masks, using the Low Resolution Imaging Spectrometer on Keck \citep{oke1995}.   We matched the H\ii\ regions based on Figure 1 of \citet{bresolin2004} which labels each targeted region on a map of NGC\,5194.  As we have adopted a slightly different optical radius ($\Delta$R$_{25}$ = 12\arcsec) we compared the resulting radial locations in Table \ref{t:bresolincomp}.  Our deprojection of NGC\,5194 yields significantly different radii for two H\ii\ regions, $-$135.4$-$181.4 and $-$86.5$-$79.4.  Given that NGC\,5194 has a nearly face-on orientation we visually inspected these locations relative to both our adopted optical R$_{25}$ and that of \citet{bresolin2004}.  We find that our adopted radii are consistent with circular apertures scaled by the appropriate radius.  Residual differences are easily accounted for by projecting the apertures using an inclination angle of 22$^\circ$ (see Table \ref{t:m51global}).  

In general, we find very good agreement between the derived abundances from \citet{bresolin2004} and those presented in this work.  In the N/O and S/O ratios, which are less sensitive to temperature, we find a mean offset of $-$0.02~dex and 0.06~dex, respectively, both of which are within the scatter of 0.10\,dex measured in each of these ratios.  Regarding the more temperature sensitive absolute oxygen abundance, on average we find slightly higher abundances (by ~0.05~dex) for these common regions.  Most abundance discrepancies (6 of 8) are well within the stated 1$\sigma$ errors and are not significant ($<$ 0.05 dex); of the two regions showing discrepancies greater than 1$\sigma$, one is consistent within two standard deviations.  In contrast, the earlier photoionization modeling of \citet{diaz1991}, with whom we have two regions in common, indicated significantly higher oxygen abundances ($\sim$0.5~dex) and correspondingly lower N/O ($\sim$0.3~dex) and S/O ($\sim$0.4~dex)  ratios.

\begin{deluxetable*}{lcccccccc}  
\tabletypesize{\scriptsize}
\tablecaption{Comparison with Previous Observations}
\tablewidth{0pt}
\tablehead{ 
  \colhead{H\ii\ Region Name}	&
  \colhead{R/R$_{25}$}	&
  \colhead{R$_{23}$\tablenotemark{a}}	&
  \colhead{[N\,\ii]/H$\beta$\tablenotemark{b}}	&
  \colhead{EW$_{H\beta}$ [\AA]}	&
  \colhead{ T[N\,\ii] [K]}	&
  \colhead{12+log(O/H)}	&
  \colhead{log(N/O)}	&
  \colhead{log(S/O)}	
  }
\startdata
NGC\,5194+71.2+135.9 	&	0.47	&	128.1	&	0.51	&	45	&	6774	&	8.58	&	-0.62	&	-1.61	\\\vspace{0.06in}
CCM 10	&	0.59	&	142.5	&	0.50	&	51	&	6900	&	8.56	&	-0.66	&	-1.54	\\
NGC\,5194+104.1-105.5 	&	0.45	&	171.6	&	0.59	&	124	&	7029	&	8.63	&	-0.65	&	-1.46	\\\vspace{0.06in}
CCM 53	&	0.51	&	174.1	&	0.54	&	92	&	6900	&	8.66	&	-0.63	&	-1.63	\\
NGC\,5194+109.9-121.4 	&	0.50	&	189.8	&	0.74	&	131	&	7389	&	8.52	&	-0.59	&	-1.35	\\\vspace{0.06in}
CCM 54	&	0.56	&	170.7	&	0.65	&	144	&	7300	&	8.49	&	-0.53	&	-1.52	\\
NGC\,5194+98.1-113.8 	&	0.46	&	114.9	&	0.48	&	88	&	6642	&	8.63	&	-0.55	&	-1.41	\\\vspace{0.06in}
CCM 55	&	0.51	&	103.7	&	0.43	&	74	&	6600	&	8.60	&	-0.49	&	-1.44	\\
NGC\,5194+83.4-133.1 	&	0.47	&	134.2	&	0.58	&	183	&	6788	&	8.60	&	-0.58	&	-1.43	\\\vspace{0.06in}
CCM 57	&	0.51	&	135.6	&	0.68	&	155	&	7400	&	8.40	&	-0.54	&	-1.50	\\
NGC\,5194-135.4-181.4 	&	0.70	&	274.6	&	0.68	&	144	&	7622	&	8.59	&	-0.80	&	-1.51	\\\vspace{0.06in}
CCM 71A	&	1.04	&	209.2	&	0.79	&	96	&	7700	&	8.45	&	-0.62	&	-1.58	\\
NGC\,5194-86.5-79.4 	&	0.37	&	63.4	&	0.31	&	51	&	6092	&	8.66	&	-0.48	&	-1.47	\\\vspace{0.06in}
CCM 72	&	0.51	&	72.0	&	0.28	&	61	&	6100	&	8.71	&	-0.63	&	-1.56	\\
NGC\,5194-4.3+63.3 	&	0.19	&	39.5	&	0.18	&	52	&	5704	&	8.84	&	-0.43	&	-1.70	\\
P203	&	0.19	&	38.6	&	0.15	&	47	&	5600	&	8.84	&	-0.46	&	-1.61	\enddata
\tablenotetext{a}{R$_{23}$ = ([O\,\ii] $\lambda$3727 + [O\,\iii] $\lambda$4959 + [O\,\iii] $\lambda$5007)/H$\beta$}
\tablenotetext{b}{[N\,\ii] $\equiv$ $\lambda$5755.}
\label{t:bresolincomp}
\tablecomments{Eight H\ii\ regions are in common between \citet{bresolin2004} and our new observations.  Results from this work are listed above the \citet{bresolin2004} results for the same H\ii\ region.  For each H\ii\ region we give the name used in this paper, NGC\,5194 followed by the offset from the galaxy center given in Table \ref{t:m51global} and the name adopted by \citet{bresolin2004}, CCM\,\# or P203.  Fluxes are given in units of H$\beta$\,=\,100.}
\end{deluxetable*}

In only one H\ii\ region, $-$159.5-116.4, we do find a significant discrepancy of 0.2\,dex.  This region is part of an H\ii\ complex in the south-eastern arm of NGC~5194 consisting of three additional H\ii\ regions that were (1) observed by both \citet{bresolin2004} and this project (+104.1-105.5, +98.1-113.8, and +83.4-133.1) and (2) have significant detections of the [N\,\ii] $\lambda$5755 line in both studies.  In each these H\ii\ regions, our measured oxygen abundance agrees with the abundance reported in \citet{bresolin2004} to within 0.03\,dex, suggesting that there are not systematic differences in our analysis.  We find that these four neighboring H\ii\ regions all have similar chemical composition (i.e., 8.5\,$<$12\,+\,log(O/H)\,$<$\,8.6) as would be expected for a ISM that is well mixed.  In contrast, \citet{bresolin2004} find an oxygen abundance of 8.40 for -159.5-116.4.

\section{Results}
\begin{figure*}[tbp] 
\epsscale{1.1}
   \centering
   \plotone{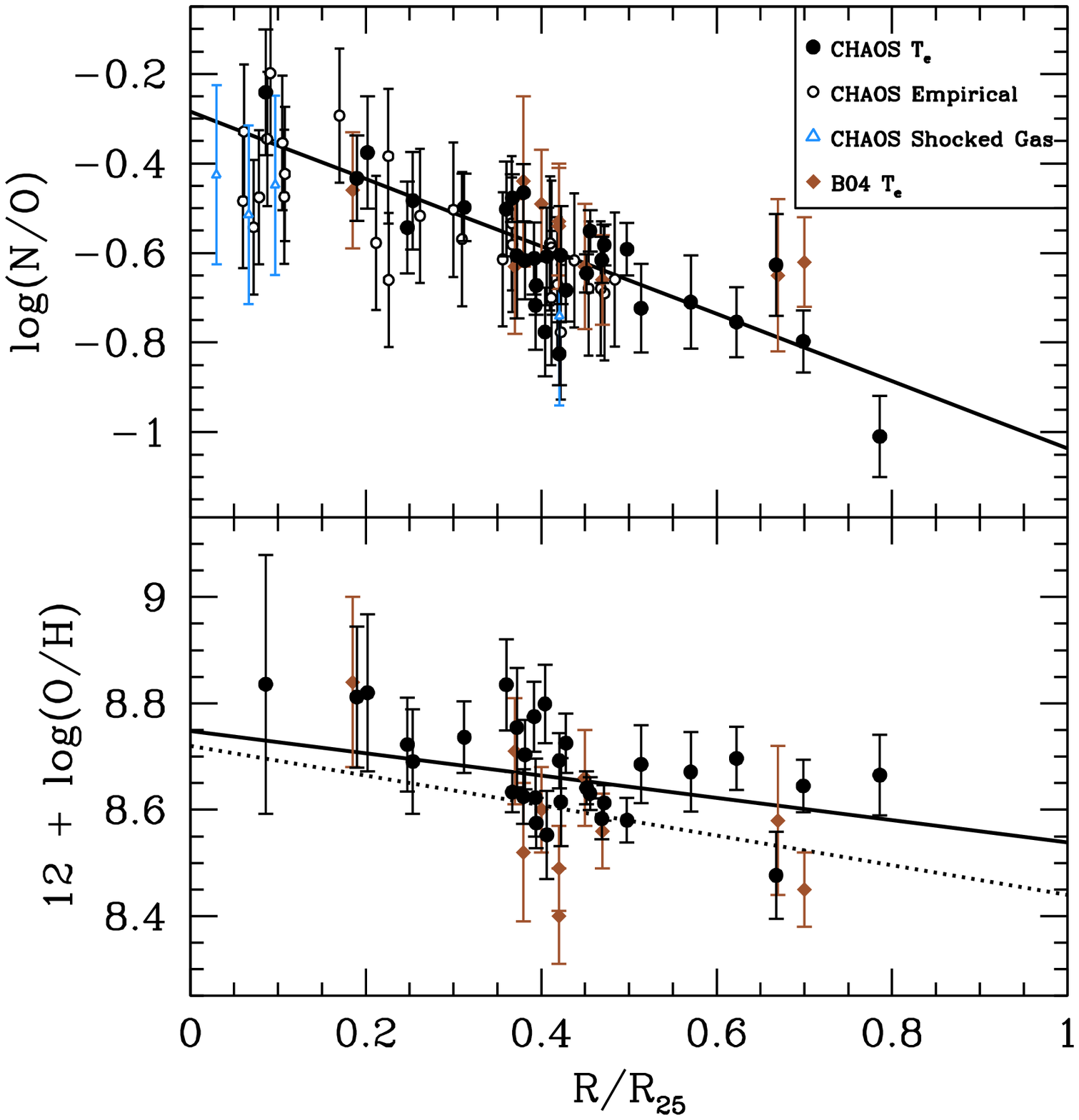}
   \caption{Radial abundance gradient in NGC~5194.  Filled points are direct abundances: black circles indicate abundances from this work; brown diamonds are direct abundances from \citet{bresolin2004}.  We plot the O/H and N/O abundance gradients we derive from direct abundance determinations as solid lines.  We plot H\ii\ regions where auroral lines are not detected on the N/O gradient, as this quantity is relatively insensitive to the electron temperature, to illustrate the extension of this trend toward the inner regions of NGC\,5194.  We denote the four regions with excessve [\ion{O}{1}] emission by blue triangles.  We also plot the gradient found by \citet{bresolin2004} as a dashed line.  For presentation, we have slightly shifted the radius of the inner most H\ii\ region from \citet{bresolin2004} so that it was visible. }  
   \label{fig:grad}
\end{figure*}  

\begin{figure*}[tbp] 
\epsscale{1.15}
   \centering
   \plotone{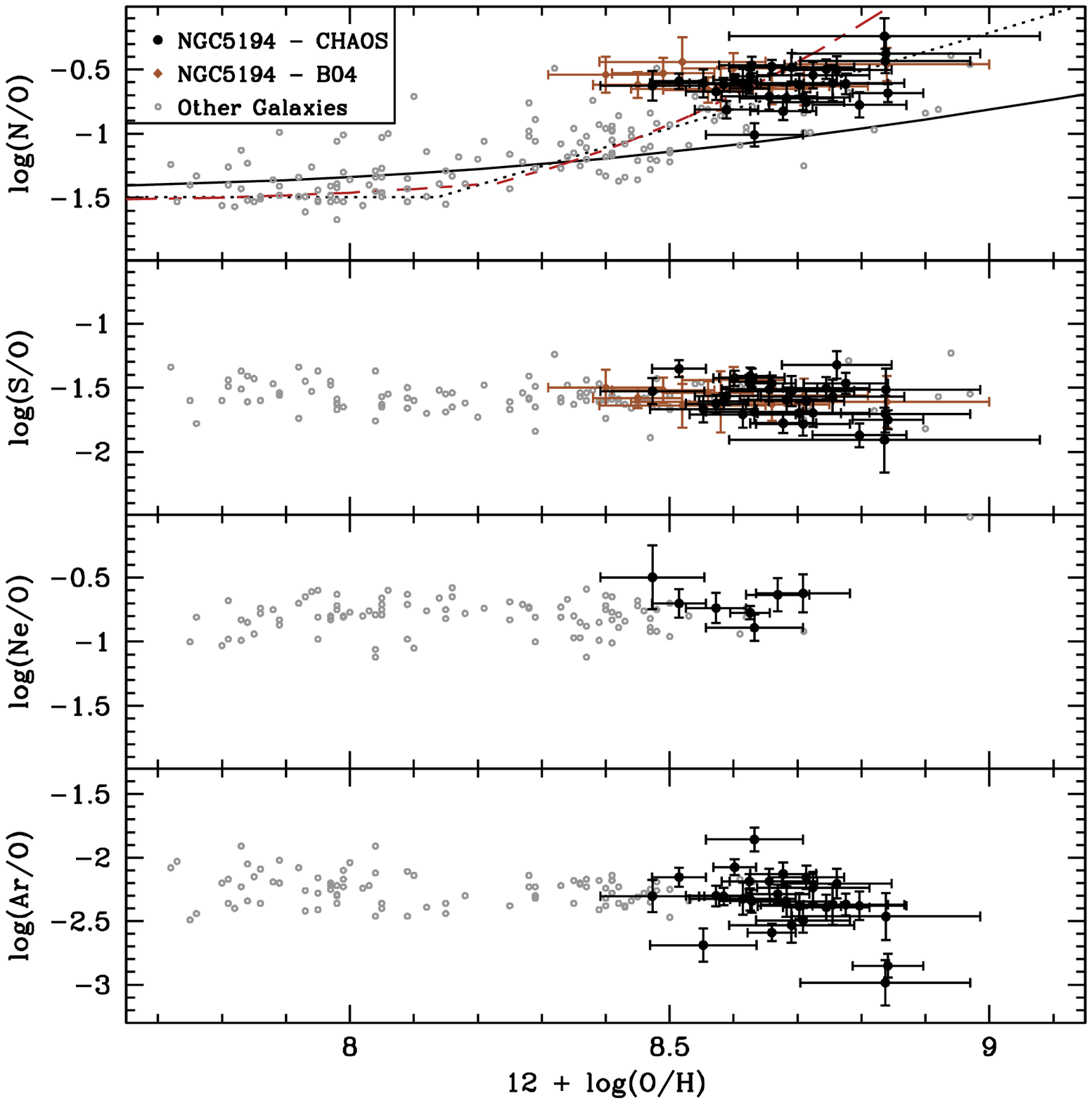}
   \caption{Relative enrichment of nitrogen and the $\alpha$ elements. Light grey circles represent direct abundances taken from \citet{bresolin2005}, \citet{bresolin2009}, \citet{bresolin2009b}, \citep{crox09}, \citet{esteban2009}, \citet{kennicutt2003}, \citet{peimbert2005}, \citet{testor2001}, \citet{testor2003}, and \citet{zurita2012};  open brown diamonds are taken from \citet{bresolin2004}. In the three lower panels, the dashed line in each panel denotes the mean value for our observations of NGC~5194. (Top -- Bottom) log (N/O) as a function of oxygen abundance. The solid line designates the theoretical curve of \citet{vilacostas}, the black dotted line is the empirical linear fit to galaxies from \citet{pilyugin2010}, the red dot-dash curve is the quadratic fit from \citet{pilyugin2010}; log (Ne/O) as a function of oxygen abundance; log (S/O) as a function of oxygen abundance; log (Ar/O) as a function of oxygen abundance.}  
   \label{fig:alpha}
\end{figure*}  

We have calculated the deprojected distance from the galaxy center in units of the isophotal radius, R$_{25}$, using the parameters given in Table \ref{t:m51global}.  We plot the radial abundance gradient of O/H and N/O in Figure \ref{fig:grad}.  Our direct abundances span the disk from R/R$_{25}\sim$0.2 to R/R$_{25}\sim$0.8, with a substantial fraction of the data concentrated in the bright star-forming arms at R/R$_{25}\sim$0.4 (17 out of 29).  We have also included direct abundances reported by \citet{bresolin2004}.  To be consistent, we have deprojected the distance to each of these H\ii\ regions using the same parameters.  We fit an error-weighted linear regression to these trends using the MPFITEX \citep{Williams2010} routine in IDL\footnote{MPFITEXY is dependent on the MPFIT package \citep{Markwardt2009}.}.  The linear regression to our direct abundance determinations are given as:
\begin{equation} 12 + \rm{log(O/H) = }
	\begin{array}{l}8.748 (\pm0.049) - 0.20 (\pm0.10)~\rm{(R/R}_{25})\\
			      8.747 (\pm0.048) - 0.0161 (\pm0.0079)~\rm{(R/kpc)} 
	\end{array}, 
\end{equation} 
 and 
\begin{equation} \rm{log(N/O) = }
	\begin{array}{l}-0.280 (\pm0.063) - 0.80 (\pm0.10)~\rm{(R/R}_{25})\\
			      -0.280 (\pm0.063) - 0.058 (\pm0.011)~\rm{(R/kpc)}
	\end{array}.
\end{equation} 
These trends are plotted as a solid lines in Figure \ref{fig:grad}.  

\section{Discussion}
\subsection{Dispersions}
The regression fits from equations (3) and (5) indicate the central O/H abundance is roughly solar or possibly slightly super-solar by 10 -- 30\%, depending on the adopted solar oxygen abundance \citep[12 + log(O/H)$_\odot$ = 8.69 -- 8.78,][]{asplund2009,ayres2013}.  The gradient hash a relatively shallow gradient with a slope of 0.02~dex/kpc, using only direct abundances, in good agreement with the gradient from \citet{bresolin2004} (see Figure \ref{fig:grad}).  Furthermore, given the dense radial coverage and the lack of a break in the N/O radial trend we can rule out the existence of a significantly steeper gradient in the region of the central bulge of NGC~5194.  

We find relatively little scatter in the O/H radial abundance trend, 0.066~dex.  We allowed the intrinsic scatter to be a free parameter and find a limit of 0.022\,dex for the intrinsic scatter in the O/H gradient, suggesting that only $\sim$30\% of the reported scatter could be intrinsic.

Similarly, we find very little scatter about the N/O gradient, 0.083~dex, with the intrinsic scatter limited to 0.042\,dex.   As the N/O ratio is both less sensitive to changes T$_{e}$ and has a larger dispersion, we interpret the these data limit the maximum possible azmuthal abundance variations to 0.042\,dex which could account for up to half the scatter.  

The scatter in the abundance gradient is well below the possible intrinsic scatter reported by \citet{berg2014} and \citet{Rosolowsky2008}, who find intrinsic dispersions of 0.13\,dex and 0.11\,dex in their studies of NGC\,628 and M33, respectively.  The analysis of \citet{berg2014} suggest that some of this dispersion may originate from temperature abnormalities that cause the [O\,\iii]\,$\lambda$4363 auroral line to be unreliable.   We find it intriguing that the dispersion is not detected in this study where the [O\,\iii]\,$\lambda$4363 is not detected.  Indeed, the only two regions where we detect this line are unusual in that the T([O\,\iii]) is quite hot relative to T([N\,\ii]) (see Table \ref{t:shocktable}).  These regions also show signs of possible shock-ionization, which is one possible mechanism suggested for these temperature abnormalities \citep{berg2014,binette2012}.    

\subsection{Gradients}
Interactions between galaxies can increase radial mixing in the ISM and thus decrease the abundance gradients.  Such shallow gradients were observed in barred galaxies by \citet{martin1994}.   This was supported by the recent work of \citet{rosa2014} and \citet{sanchez2014} who found evidence that merging and interacting systems exhibit shallow gradients compared to isolated spiral galaxies such as NGC\,300 \citep[$-$0.43$\pm$0.06\,dex/R/R$_{25}$][]{bresolin2009} and 
NGC\,5457 \citep[M101, $-$1.02$\pm$0.10\,dex/R/R$_{0}$][]{kennicutt2003}.  Compared to these systems, we find NGC\,5194 also exhibits a shallow gradient ($-$0.20 $\pm$0.10\,dex/R/R$_{25}$).  This may indeed result from interactions of NGC~5194 with its nearby companion NGC5195 as metal-enriched gas could have been mixed into the disk of NGC~5194 during the recent interaction which also served to ignite star formation \citep[e.g.,][]{ellison2013}.  Our derived slope in the O/H gradient of $-0.20$ dex/(R/R$_{25}$) falls in the parameter space occupied by interacting systems \citep{rosa2014}.  One may expect that gas in the outskirts of NGC~5194 would be more significantly affected by dynamical interactions.  While we do not detect any flattening of the abundance gradients in the outskirts of the disk of NGC~5194, we do not consider this to be highly constraining as our observations only extend out to R/R$_{25}~\approx$ 0.8, well within the optical radius.

\subsection{Relative abundances and Chemical Evolution}
We plot N/O, S/O, Ne/O, and Ar/O as a function of O/H in Figure \ref{fig:alpha}.  The S/O, Ar/O, and Ne/O ratios are consistent with the constant values observed in massive and low mass star-forming galaxies \citep{crox09,bresolin2005}.  We note that the most metal rich H\ii\ regions (12 + log(O/H) $\gtrsim$ 8.75) appear to exhibit slightly depressed Ar/O and S/O ratios.  Rather than indicating a real change in the relative abundances of these $\alpha$-elements, this is most likely a systematic effect caused by complexities of ionization correction factors in low excitation H\ii\ regions \citep{stasinska2001}.  This conclusion is strengthened by the fact that we find both (S$^+$ + S$^+$$^+$)/(O$^+$ + O$^+$$^+$) and (Ar$^+$$^+$)/(S$^+$$^+$) are low in these cool H\ii\ regions. We confirm that the N/O is slightly elevated compared to other metal-rich galaxies, consistent with increased secondary production of nitrogen at higher O/H.  This overabundance of nitrogen relative to oxygen has been attributed to differing star formation histories \citep[e.g.,][]{henry2000}.

Similar to the findings of other studies of spiral galaxies \citep[e.g.,][]{kennicutt2003}, but in contrast with \citet{bresolin2004}, we find a steeper gradient in the N/O ratio, with a slope of 0.057~dex/kpc using only direct abundances and 0.05~dex/kpc using all H\ii\ regions.   While N/O is higher than the expected theoretical relation of \citet{vilacostas}, it is consistent with empirical predictions of N/O based on other nearby spiral galaxies, see Figure \ref{fig:alpha}.  This could also be evidence of a much steeper rise in N/O as a function of oxygen abundance \citep{henry2000} relative to the predictions of \citet{vilacostas}.  Although the observed excess in N/O is occurring at a different metallicity than predicted, this is likely due to the adopted metallicity scale.  Indeed, if we adopt calibrations based purely on photoionization models \citep{KK04}, rather than direct temperatures, we obtain higher oxygen abundances ($<$12~+~log(O/H)$_{KK04}>$~=~9.17), with a more compressed scale similar to what is seen by \citet{henry2000}.

Another possibility is that massive Wolf-Rayet stars have selectively enhanced the nitrogen present in the ISM of NGC~5194 \citep{kobulnicky1997}.  In Table \ref{t:locations}, we note regions where in the broad Wolf-Rayet features located near $\lambda$4660\,\AA\ have been detected.  Although significant detections of these broad features are noted in only 12 apertures, hints of these features appear in several additional apertures.  This indicates the presence of large numbers of these objects, even if only in small numbers, typically 1--2, for a given H\ii\ region \citep{Schaerer1998}.  Regardless of the mechanism, it is clear that secondary production of nitrogen is dominant in NGC~5194 as the average N/O ratio is larger than the nominal value of primary production (log(N/O) $\sim-$1.5) by a factor of 7.5.

Our measurements of the S/O ratio do not yield a significant gradient.  However, \citet{diaz1991} found evidence for a central depression in the S/O ratio in NGC~5194.  While the innermost H\ii\ regions exhibit somewhat depressed S/O (see Figure \ref{fig:alpha}), this is likely a result of underestimating the ionization correction factor for sulfur due to the break down of the assumed temperature structure in these very low ionization nebula.  Even though the N/O ratio is insensitive to changes in temperature, the same is not necessarily true of S/O ratio with relies upon an ionization correction factor that depends on O$^+$/O \citep{garnett1989}.  Indeed, comparing the ion ratio (S$^+$ + S$^{++}$)/(O$^+$ + O$^{++}$) to the ionization correction factor indicates that these cool regions have significantly underestimated the necessary correction factor as most of the power in [S\,\iii] and [O~\iii] lines has shifted to the infrared where lines are (a) not measured and (b) insensitive to temperature \citep{crox13}.  Furthermore, the N/O and O/H gradients do not exhibit a similar central depression.

\section {Conclusions}
To gain a greater understanding of the chemical composition of spirals we have undertaken CHAOS, a spectroscopic study of several of the best studied nearby spiral galaxies.  Here we present data on one of these galaxies, NGC~5194, for which we have obtained high signal-to-noise spectra for 59 H\ii\ complexes.  Out of these, 28 distinct H\ii\ regions have detections of the temperature-sensitive auroral lines [N\,\ii] $\lambda$5755 or [S\,\iii] $\lambda$6312, permitting the derivation of a direct oxygen abundance. Using semi-empirical methods, we determine N/O ratios for all 59 regions.

We confirm the findings of \citet{bresolin2004} that NGC~5194 has a roughly solar metallicity and a relatively flat metallicity gradient.  We present a robust metallicity gradient with relatively small scatter between 0.06 $<$ R/R$_{25}$ $<$ 0.11.  As temperature-sensitive lines are not detected in the innermost part of NGC~5194, we co-add spectra and measure a direct oxygen abundance to confirm the adopted linear abundance gradient.

We also establish a robust gradient in the N/O ratio that is significantly steeper than the gradient in O/H.  This points to significantly enhanced secondary nitrogen production, $\sim$7.5 times more than primary nitrogen production.  In contrast we find no significant changes in Ne/O, Ar/O, nor S/O across the disk of NGC~5194, or even in comparison to metal poor dwarf galaxies \citep{crox09}.  Rather, line ratios suggest that the ionization correction factors typically employed for S and Ar may become unreliable at very cool, T$\sim$4500~K, temperatures.

As spiral galaxies are dominant sites of star formation in the Universe, significant effort has been given to understanding them and their evolution.  One major aspect of that evolutionary process is the chemical evolution that occurs in the interstellar medium of these galaxies.  However, most of our knowledge of the chemical make up of massive galaxies can be summarily stated as metallicity correlates with mass \citep{tremonti2004} and spiral galaxies exhibit radial gradients \citep{pageledmunds1981}.  While this description of present day spiral galaxies is secure, it is only a first-order description of chemical evolution.  In order to understand the evolution of galaxies we must obtain a more complete understanding of chemical enrichment.  Here we highlight the data being collected by the CHAOS study for NGC\,5194.  With similar high quality spectra forthcoming for the rest of the sample, we will be able to establish detailed chemical abundance patterns for several nearby galaxies and greatly improve our understanding of chemical enrichment.

~\\
\acknowledgments
K.V.C. is grateful for support from NSF Grant AST-6009233.  
This paper uses data taken with the MODS spectrographs built with funding from NSF grant AST-9987045 and the NSF Telescope System Instrumentation Program (TSIP), with additional funds from the Ohio Board of Regents and the Ohio State University Office of Research.  
This paper made use of the modsIDL spectral data reduction pipeline developed in part with funds provided by NSF Grant AST-1108693.  
We are grateful to D. Fanning, J.\,X. Prochaska, J. Hennawi, C. Markwardt, and M. Williams, and others who have developed the IDL libraries of which we have made use: coyote graphics, XIDL, idlutils, MPFIT, MPFITXY, and impro.  
This work was based in part on observations made with the Large Binocular Telescope (LBT). The LBT is an international collaboration among institutions in the United States, Italy and Germany. The LBT Corporation partners are: the University of Arizona on behalf of the Arizona university system; the Istituto Nazionale di Astrofisica, Italy; the LBT Beteiligungsgesellschaft, Germany, representing the Max Planck Society, the Astrophysical Institute Potsdam, and Heidelberg University; the Ohio State University; and the Research Corporation, on behalf of the University of Notre Dame, the University of Minnesota, and the University of Virginia.
This research has made use of the NASA/IPAC Extragalactic Database (NED) which is operated by the Jet Propulsion Laboratory, California Institute of Technology, under contract with the National Aeronautics and Space Administration.

\end{document}